%% file: 6227.tex
\documentstyle[psfig]{l-aa}
\include{alias}

\begin{document}
\topmargin=1.5cm
\thesaurus{ 09.13.2;  
            09.03.1;  
            11.02.2;  
	    11.09.4;  
            11.17.1   
            13.18.1;  
	         }
   \title{Molecular absorption lines at high redshift:}
   \subtitle{PKS1413+135 (z=0.247)}
%
   \author{T.~Wiklind\inst{1}, F.~Combes\inst{2}}
   \offprints{T.~Wiklind, tommy@oso.chalmers.se}
   \institute{
              Onsala Space Observatory, S--43992 Onsala, Sweden
   \and
              DEMIRM, Observatoire de Paris, 61 Av. de l'Observatoire,
              F--75014 Paris, France
             }
   \date{Received date; Accepted date}
   \maketitle
   \markboth{Wiklind \& Combes}
            {Molecular absorption lines at high redshift}
%
\begin{abstract}

A detailed study of the absorbing molecular clouds towards the 
radio source PKS1413+135, at a redshift z$=$0.247, is reported.
Physical conditions (density, temperature, filling factors)
are derived for the molecular gas, in the frame work of two
models: a homogeneous multicomponent model with equal filling
factors, and a two--phase medium consisting of dense clumps
embedded in a more diffuse component.
It is shown that our absorption data are consistent with the
presence of a diffuse gas component, dominating the observed
opacity, and a dense component, accounting for most of the mass.
We also show that without knowledge of the small scale structure
of the absorbing molecular gas, we can only derive lower limits
to the column density.
Given the very narrow absorption spectrum, the size of the overall
absorbing cloud along the line of sight must be quite small, of the
order of 1\,pc.
The variability of the absorption spectrum has been studied over a
time range of more than 2 years.
We find that the opacity ratio between two absorbing components
in CO has varied by a factor $2.3 \pm 0.3$.
The variations are interpreted as a change of the line of sight
due to structural changes in the background source.
We discuss what can be derived from molecular absorption line
observations concerning the invariance of physical constants,
such as the mass of the molecules. We discuss molecular line
data from redshifts 0.27--0.89, corresponding to look--back times
of 30\% to 60\% of the age of the universe.

\end{abstract}
\section{Introduction}

Studies of absorption lines towards high redshift QSOs
give detailed information about the interstellar medium
(ISM) in distant objects at a linear resolution limited
only by the size of the background continuum source.
Absorption line observations also have a superior
sensitivity compared to the corresponding emission
lines and can successfully be used as probes of the
composition as well as the physical and chemical
conditions in the ISM.
Such observations have for the most part been done
at optical wavelengths, but the last few years have
seen an increase in the number of high redshift
absorption line systems observed at radio wavelengths.

The neutral atomic gas component has been observed
through absorption of the 21\,cm\ line of atomic
hydrogen. About a dozen 21\,cm HI absorbers at
redshifts $z \ga 0.2$ have been detected (Carilli 1995),
with typical column densities
$N(\hI)=5 \times 10^{18}\,(T_{\rm sp}/f_{\rm HI})$\,\cmsq\ 
(e.g. Briggs 1988; Carilli 1995), where $T_{\rm sp}$
is the spin temperature in K and $f_{\rm HI}$ is the
covering factor of atomic gas across the extent of the
background continuum source.

Compared to the ionized and atomic parts of the ISM,
the molecular gas component is characterized by high
densities and low temperatures. Stars are formed directly
from the molecular gas, and its physical and chemical
status gives information about the stellar formation
history as well as giving the initial conditions for
the on--going star formation.
We have recently detected absorption of molecular
rotational transitions in four galaxies at redshifts
0.25--0.89 (Wiklind \& Combes 1994, 1995, 1996a,b;
Combes \& Wiklind 1996). The background source is in
all cases a highly obscured radio--loud QSO, located
either in the background or in the galaxy where the
absorption takes place.
Absorptions in the millimeter wave band provide a velocity
resolution more than two orders of magnitude higher
than in optical spectroscopy.
Since the amount of molecular gas needed to produce a
detectable absorption line is less than 1\,\mo, the
sensitivity of molecular absorption line observations in
terms of total mass is $\sim 10^{12}$ times higher than
for the corresponding emission lines (cf. Wiklind \&
Combes 1994). This permits us to derive physical properties
of the molecular component of the ISM at high redshifts
which would otherwise be impossible.

\medskip

In this paper we present new molecular absorption line
data on the absorption system at z$=$0.24671 towards
PKS1413+135. In Sect.\,2 we give a general background to
the radio source PKS1413+135 and the absorbing galaxy.
The observations and data reduction are presented in
Sect.\,3. Our results are given in Sect.\,4:
in particular, we derive the physical properties of the
molecular gas and we discuss possible time variations in
the absorption lines. In Sect.\,5 we compare the chemistry
of the redshifted molecular gas with Galactic and nearby
extragalactic systems and we discuss constraints to
variations in physical constants derived from our absorption
line data. Finally in Sect.\,6 we summarize our results.

\section{PKS1413+135 and the galaxy at z$=$0.247}

The radio source PKS1413+135 has a flat continuum spectrum
extending from a few hundred MHz into the near--infrared,
with an exponential decrease at $\sim$5\,$\mu$m (e.g. McHardy
et al. 1991).
The radio loud AGN is invisible at optical wavelengths.
On the basis of this steep infrared to optical continuum,
PKS1413+135 was originally classified as a red QSO
(e.g. Bregman et al. 1981).
The abrupt decrease of the continuum in the near--infrared
was modeled as a high--energy cut--off in the electron
energy distribution. Spectroscopic surveys in both the
infrared and optical bands showed no prominent emission
lines (Bregman et al. 1981; McHardy et al. 1994; Perlman
et al. 1996).
Stocke et al. (1992) detected Balmer absorption
lines and weak [OII] 3727\,\AA\ emission at z$=$0.247.
Rapid variability at radio and infrared wavelengths, strong
and rapid variable infrared polarization and a flat--spectrum
radio continuum, lead to a classification of PKS1413+135
as a BL Lac object. Subsequent radio imaging using VLBI
techniques has questioned this classification. The radio
source consists of: (1) a compact radio core, (2) a jet--like
structure on a parsec scale, and (3) a counter--jet
(Perlman et al. 1996). All the features except the compact core
have steep spectral indices, meaning that their contribution to
the continuum flux is neglible at millimeter wavelengths.
The presence of a counter--jet disagrees with the interpretation
of PKS1413+135 as a BL Lac object, since these are believed to
represent radio sources with the jet pointing almost exactly
along the line--of--sight, with the flux boosted by relativistic
beaming. The counter--jet would not be visible in this scenario.
Perlman et al. (1996) model PKS1413+135 as a typical Compact
Symmetric Object (CSO); a young radio source with an age
$\leq 10^4$\,yrs. Doppler--boosting of the continuum could
occur in the core, but not in the parsec scale jet and
counter--jet.

Imaging of the system in optical and infrared bands reveals
an edge--on galaxy (inclination $i \approx 87^{\circ}$) with
a prominent dust lane (McHardy et al. 1991, 1994; Stocke et al.
1992). The luminosity profile is exponential with a scale
length of 1.8'', corresponding to 5.8\,kpc for
H$_0 = 75$\,\kmsmpc\ and q$_0 = 0.5$ (these values are used
throughout this paper).
The disk--to--bulge ratio is $\sim$1.1 (McHardy et al. 1994),
but this value is uncertain due to unknown extinction
corrections. The integrated blue luminosity (obtained from the
V--band which approximately corresponds to the B--band in the
restframe of the galaxy) is $19.6 \pm 0.15$ (McHardy et al. 1992),
giving $M_{\rm B} = -20.49$ or $L_{\rm B} = 2.4 \times 10^{10}$\,\lo.
No sign of the AGN is seen at optical wavelengths, but in
the infrared the luminosity becomes dominated by a point source.
This point source is situated within 0.1'' (325\,pc) of the
center of the galaxy (Stocke et al. 1992). If the radio
source is situated behind the galaxy, this close alignment
would lead to gravitational lensing. No signs are found for
image multiplicity or distortion, in either optical or
radio bands (e.g. Stocke et al. 1992; Perlman 1996), strongly
suggesting that the galaxy at z$=$0.247 is host to the radio source.

The X--ray spectrum shows a deficiency of low--energy X--ray
photons, indicative of an extinction $A_V > 30$\,mag (Stocke et al.
1992; McHardy et al. 1994). This corresponds to an absorbing
screen with N(H)$>10^{22}$\,cm$^{-2}$. Carilli et al. (1992)
detected redshifted 21\,cm absorption at z$=$0.24671, with
a column density N(HI)$=1.3 \times 10^{19}(T_{\rm s}/f_{HI})$,
where $T_{\rm s}$ is the spin--temperature and $f_{HI}$ is the
covering factor of atomic gas across the background source.
Wiklind \& Combes (1994) reported molecular absorption at the
redshift of the galaxy, with N(H$_2$)$=4 \times 10^{20}$\,cm$^{-2}$.
These column densities are much smaller than that inferred from
the X--ray data, but suggest that the galaxy at z$=$0.247 is
of late type. The presence of molecular and atomic gas, the
exponential luminosity profile and the broad--band colours
are consistent with the galaxy being of morphological type
Sb--c.

Hence, PKS1413+135 is a radio source situated in a spiral galaxy
at a redshift z$=$0.24671. The blue luminosity of the galaxy is
typical for an Sbc. The abrupt cut--off in the optical
continuum is caused by extinction, and not by a cut--off in the
electron energy distribution.
An excess in near--infrared emission should be present due to
dust heated by radiation from the AGN or by the interaction
of the relativistic electrons with the ambient interstellar
medium.
No such excess has been observed (cf. Stocke et al. 1992).

\begin{table}
\begin{flushleft}
\caption[]{Observed molecules and transitions}
\scriptsize
\begin{tabular}{lcr|rcc}
\hline
 & & & & & \\
%
\multicolumn{1}{c}{Mol.}                     &
\multicolumn{1}{c}{Transition}                   &
\multicolumn{1}{c|}{$\nu_{\rm rest}^{a)}$}       &
\multicolumn{1}{c}{$\nu_{\rm obs}^{b)}$}         &
\multicolumn{1}{c}{Det.$^{c)}$}                  &
\multicolumn{1}{c}{Tel.$^{d)}$}                  \\
\multicolumn{2}{c}{ }                            &
\multicolumn{1}{c|}{GHz}                         &
\multicolumn{1}{c}{GHz}                          &
\multicolumn{2}{c}{ }                            \\
 & & & & & \\
\hline
 & & & & & \\
CO             & 0$\rightarrow$1 & 115.271204 &  92.460319 & Yes & I \\
 & & & & & \\
$^{13}$CO      & 0$\rightarrow$1 & 110.201370 &  88.393748 & No  & I \\
 & & & & & \\
\hline
 & & & & & \\
HCN            & 0$\rightarrow$1 &  88.631847 &  71.092593 &  No & K \\
               & 1$\rightarrow$2 & 177.261111 & 142.183115 & Yes & I \\
               & 2$\rightarrow$3 & 265.886180 & 213.270271 & Yes & I \\
 & & & & & \\
\hline
 & & & & & \\
HCO$^+$        & 0$\rightarrow$1 &  89.188523 &  71.539109 & Yes & K \\
               & 1$\rightarrow$2 & 178.375065 & 143.076630 & Yes & I \\
               & 2$\rightarrow$3 & 267.557619 & 214.610951 & Yes & I \\
 & & & & & \\
H$^{13}$CO$^+$ & 1$\rightarrow$2 & 173.506782 & 139.171726 & No  & I \\
 & & & & & \\
\hline
 & & & & & \\
HNC            & 1$\rightarrow$2 & 181.324758 & 145.442611 & Yes & I \\
               & 2$\rightarrow$3 & 271.981067 & 218.159048 & Yes & I \\
 & & & & & \\
\hline
 & & & & & \\
N$_2$H$^+$     & 0$\rightarrow$1 &  93.173400 &  74.735424 & No  & K \\
               & 1$\rightarrow$2 & 186.344874 & 149.469302 & No  & I \\
               & 2$\rightarrow$3 & 279.511671 & 224.199430 & No  & I \\
 & & & & & \\
\hline
 & & & & & \\
CN             & 0$\rightarrow$1 & 113.490980 &  91.0323830 & Yes  & I \\
 & & & & & \\
\hline
 & & & & & \\
CS             & 1$\rightarrow$2 &  97.980950 &  78.591613 & No  & K \\
 & & & & & \\
\hline
 & & & & & \\
H$_2$CO  & $1_{1,1}\rightarrow 2_{1,2}$ & 140.839504    & 112.968937     & No  & I \\
 & & & & & \\
\hline
 & & & & & \\
O$_2$          & $1_{0}\rightarrow 1_{1}$ & 118.750    & 95.251     & No  & I \\
 & & & & & \\
\hline
\end{tabular}
\ \\
a)\ The rest--frequency of the observed molecules taken from Lovas (1992). \\
b)\ Derived from $\nu_{\rm obs} = \nu_{\rm rest}/(1+z_{\rm a})$. \\
c)\ Detection: Yes/No. \\
d)\ Telescope: I$=$IRAM, K$=$Kitt Peak 12--m.
\end{flushleft}
\end{table}

\begin{figure*}
\psfig{figure=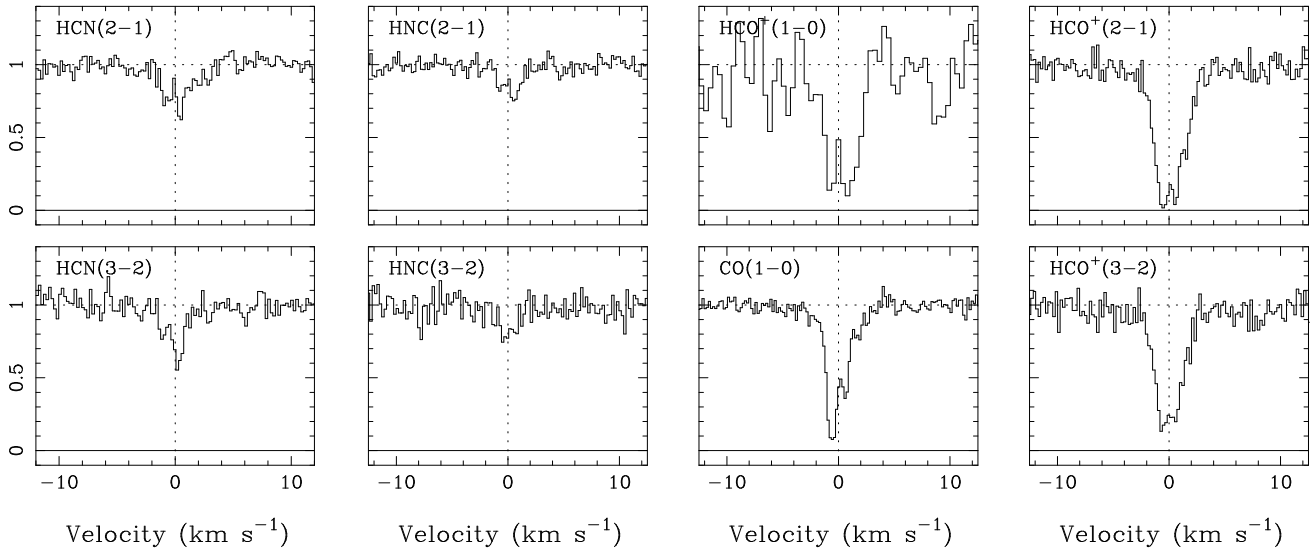,bbllx=28mm,bblly=24mm,bburx=105mm,bbury=200mm,width=17.5cm,angle=-90}
\caption[]{Total averages of the J$=$0$\rightarrow$1 transitions
of CO and HCO$^+$, the J$=$1$\rightarrow$2 and J$=$2$\rightarrow$3
transitions of HCN, HCO$^+$ and HNC towards PKS1413+135. The
velocity resolutions are $\sim$0.2\,km\,s$^{-1}$ for all spectra.
All transitions have been obtained with the IRAM 30--m telescope,
except HCO$^+$(0$\rightarrow$1) which was observed with the NRAO 12--m
telescope at Kitt Peak. The continuum levels have been normalized to unity and
zero velocity corresponds to a heliocentric redshift of 0.24671.
}
\end{figure*}

\section{Observations and data reduction}

\subsection{Observations}

The observations presented in this paper have mainly been
obtained at the IRAM (Institut de Radio Astronomie Millimetrique)
30--m telescope on Pico Veleta in Spain during several
observing runs from February 1994 to July 1996. Some
absorption lines were also observed at the NRAO
12--m telescope at Kitt Peak\footnote{NRAO is a facility of
the National Science Foundation, operated under cooperative agreement
by Associated Universities Inc.} in February 1996.

At the IRAM 30--m telescope we used 3--, 2-- and 1.3--mm
SIS receivers, tuned to the redshifted frequencies of the
molecular transitions in question (see Table\,1).
The observations were done with a nutating subreflector,
switching symmetrically $\pm$90\asec\ in azimuth with a
frequency of 0.5\,Hz. The continuum levels of the observed
sources were determined using a continuum backend and 
increasing the subreflector switch frequency to 2\,Hz.
The image sideband rejection of the receivers varied
from 0\,dB (double sideband) to 20\,dB (single sideband),
and was taken into account to determine the continuum
temperatures $T_{\rm c}$ displayed in Table\,2.
The typical system temperature ranged from 200\,K and
up, depending on the whether conditions. Data obtained
under bad weather conditions have been discarded.
For the line observations we used broad filterbanks and
narrow band autocorrelators, the former giving velocity
resolutions 1--3\,\kms, depending on the observed
frequency, and a bandwidth of 500--1000\,\kms.
With the autocorrelators we chosed frequency resolutions
of 10\,kHz for the 3--mm receiver, 20\,kHz for the 2--mm
and 40\,kHz for the 1--mm receivers. The corresponding
velocity resolutions are 30\,\ms, 40\,\ms\ and 50\,\ms,
respectively. The bandwidths for the autocorrelators were
40--50\,\kms.
The pointing of the telescope was checked regularly on
the observed sources or on nearby continuum sources.
Since the weather was good and stable for most of the
observations, we could monitor the continuum flux of
the observed source during the integration to directly
check the pointing of the telescope. Typical pointing
corrections were 5--10''.

At the NRAO 12--m telescope we used the SIS receiver
for the lower part of the 3--mm band, tuned to the
redshifted frequencies of the observed transitions.
The receiver contains two mixers allowing dual
polarization observations. For the lowest frequencies,
only one of the mixers could be used due to high
receiver temperatures in the other. Typical system
temperatures were in the range 300--500\,K.
The observations were done with a dual chopping of
the subreflector; the chopping was done at a rate
of 1.25\,Hz with a beamthrow of 2'. As backends
we used the filterbanks, giving spectral resolutions
of 0.25 and 0.50\,MHz. The pointing and focus were
checked regularly on Venus and Saturn.

The data presented here is given in chopper--wheel
calibrated antenna temperatures (\tastar). The NRAO
data is initially represented in the $T_{\rm R}^{*}$
scale. To transform it into the $T_{\rm A}^{*}$
temperature scale we multiplied the $T_{\rm R}^{*}$
values with the forward spillover and scattering
efficiency, $\eta_{fss} = 0.68$.

\subsection{Data reduction}

Each spectrum is the average of several 5--10
minutes integrations with a calibration in--between. 
Individual spectra were inspected for interference
and bad spectral baselines. The channel--to--channel noise
rms was used as a weighting factor when adding
individual spectra together. Usually no baseline
corrections were applied, although a small set
of spectra obtained under less favourable weather
conditions were corrected with a 3rd order polynomial.
The continuum level was derived directly from the
spectra, taking care to only include data obtained
under stable weather conditions. At the IRAM telescope
we also used a continuum backend and a fast chopping
technique to derive $T_{\rm c}$. This was done at
all the observed frequencies. The accuracy of the
continuum level is critical for the subsequent
interpretation of the data (see below).
The spectra have not been binned or smoothed except
in the cases when adding together data obtained
with different frequency resolutions. In these
cases spectra are rebinned to correspond to 
the spectra with the lowest spectral resolution.
The spectra presented here have their continuum
level normalized to unity.

\subsection{The continuum level}

The observed continuum level $T_{\rm c}$ of the
background source is an important quantity when
deriving parameters from the absorption lines.
It is, however, difficult to obtain $T_{\rm c}$
with an accuracy better than 10\% in a single
measurement, due to calibration uncertainties
inherent in the atmospheric models used for the
chopper--wheel method. The variance about the mean
of the opacity  can be obtained by differentiating
Eq.\,(4) in Sect.\,4.2 and is
\begin{equation}
\sigma_{\tau_{\nu}} \approx
{\exp(\tau_{\nu}) - 1 \over T_{\rm c}}
\sigma_{\rm T_{\rm c}} \ ,
\end{equation}
where $\sigma_{\rm T_{\rm c}}$ is the standard deviation
of the observed continuum temperature and we have
neglected the variance of $T_{\rm abs}$, the measured depth of the
absorption line, which generally is much smaller than
that of the continuum temperature.
For $\tau << 1$, $\sigma_{\tau_{\nu}}/\tau \approx 
\sigma_{\rm T_{\rm c}}/T_{\rm c}$,
which is usually small for finite values of $\tau$ and $T_{\rm c}$.
For $\tau >> 1$, $\sigma_{\tau_{\nu}} \approx \exp(\tau)
\sigma_{\rm T_{\rm c}}/T_{\rm c}$ and
we loose all information about the true opacity.
In the critical opacity region ($\tau \ga 1$),
the error in our estimate of the opacity depends
nonlinearly on the continuum level and the opacity
itself.

More specifically, the important parameter when
deriving the opacity of an absorption line seen
towards an unresolved background source, is the
line--to--continuum ratio.
The values derived are therefore independent of
the true value of the continuum level as long
as the only variations in $T_{\rm c}$ originate
in  instantaneous pointing uncertainties. This
is the situation under good weather conditions,
where the average calibration uncertainty is the
only parameter affecting the derived opacity.
When the atmospheric conditions change between two
calibrations (typically 5--10 minutes), we loose
information about the continuum level and the
line--to--continuum ratio ceases to be a good
measure of the opacity. When this has been the
case, we chose a period of acceptable atmospheric
conditions and adopted the derived $T_{\rm c}$ from
this period to scale the rest of the data.
However, in this procedure we become sensitive
to pointing errors. Fortunately, their variance
was always lower than 20\% of the beam--width.

The main uncertainty in the derived column densities,
excitation temperatures and, to a lesser extent,
abundance ratios, originates in the uncertainty
of the continuum level. The nonlinear dependence
makes it difficult to quantify the formal uncertainty.
Formal errors presented here are therefore based on
the assumption that $T_{\rm c}$ is correct.
The relative errors in $T_{\rm c}$ are then to be
multiplied by 1.3, 1.7 and 3.2 to obtain the relative
errors induced on $\tau$, when 
$\tau_{\nu}$ = 0.5, 1 and 2 respectively.

\section{Results}

\subsection{The observed absorption lines}

Towards the radio source PKS1413+135 we have observed 18
different molecular rotational transitions of a total
of 11 molecules. Of these we clearly detect 9
transitions of 5 different molecules; the J$=$0$\rightarrow$1
lines of CO, HCO$^+$ and CN\footnote{For CN we clearly
detect one hyperfine transition: N$=$1--0, J$=$3/2--1/2,
F$=$5/2--3/2.}, the J$=$1$\rightarrow$2 and
J$=$2$\rightarrow$3 lines of HCN, HCO$^+$ and HNC.
The observed transitions with their rest--frequencies
and corresponding redshifted frequencies are presented
in Table\,1. In Table\,2 we list the observed integrated
optical depths and the derived excitation temperatures
and column densities. Notice that some of the column
densities in Table\,2 are modified after a more thorough
analysis (Sect.\,4.2.3). The final column densities are
presented separately in Table\,3.

The detected transitions are shown in Figs.\,1 and 2.
The continuum levels have been normalized to unity.
All the spectra in Fig.\,1 except HCO$^+$(0$\rightarrow$1)
are averages of the observing sessions in 1994 (cf. Sect.\,3).
The 1995 and 1996 data have not been used since there might
be temporal changes in the absorption profiles (see Sect.\,4.5).
In order to show the weak CN(0$\rightarrow$1) line, the
scaling in Fig.\,2 is different from that of Fig.\,1.
The CN hyperfine transitions which fall within the spectrometer
are the J$=3/2-1/2$ with F$=1/2-3/2$, F$=3/2-3/2$, F$=1/2-1/2$,
F$=5/2-3/2$ and F$=3/2-1/2$.
Only the F$=5/2-3/2$ line is formally detected, possibly
blended with the F$=3/2-1/2$ line. The height of the markings
corresponds to the observed intensities in the Orion hot core
(cf. Lovas 1992).

Some of our non--detected transitions are shown in Fig.\,3.
Since the isotopic lines of $^{13}$CO(0$\rightarrow$1) and
H$^{13}$CO$^+$(1$\rightarrow$2) remain undetected at low
noise levels, the corresponding $^{12}$CO and H$^{12}$CO$^+$
lines are not saturated, despite absorption depths close
to unity. This means that the derived column densities are
not affected by saturation effects. The peak opacities
for the CO(0$\rightarrow$1) and HCO$^+$(1$\rightarrow$2)
lines are 2--3 (see Table\,2).

The column density of CO is estimated to be approximately
$2 \times 10^{16}$\,cm$^{-2}$. With a Galactic CO to H$_2$
abundance ratio of $5 \times 10^{-5}$ (e.g. van Dishoeck \&
Black 1987), the column density of molecular hydrogen towards
PKS1413+135 at z$=$0.24671 is $\sim 4 \times 10^{20}$\,cm$^{-2}$.

\begin{table*}
\begin{flushleft}
\caption[]{Observed properties for PKS1413+135}
\tiny
\begin{tabular}{lcrcccclrr}
\hline
 & & & & & & & & \\
\multicolumn{1}{l}{Molecule}                     &
\multicolumn{1}{l}{Transition}                   &
\multicolumn{1}{c}{$T_{\rm c}^{a)}$}             &
\multicolumn{1}{c}{$\int{\tau_{\nu} dV}$}        &
\multicolumn{1}{c}{$\tau_{0}^{b)}$}              &
\multicolumn{1}{c}{$\sigma_{\tau}^{c)}$}         &
\multicolumn{1}{c}{$\delta V^{d}$}               &
\multicolumn{1}{c}{Date(T)}                      &
\multicolumn{1}{c}{$T_{\rm x}^{e)}$}             &
\multicolumn{1}{c}{$N$}                          \\
 & & & & & & & & \\
\multicolumn{1}{c}{ }                            &
\multicolumn{1}{c}{ }                            &
\multicolumn{1}{c}{mK}                           &
\multicolumn{1}{c}{km\,s$^{-1}$}                 &
\multicolumn{1}{c}{ }                            &
\multicolumn{1}{c}{km\,s$^{-1}$}                 &
\multicolumn{1}{c}{km\,s$^{-1}$}                 &
\multicolumn{1}{c}{ }                            &
\multicolumn{1}{c}{K}                            &
\multicolumn{1}{c}{cm$^{-2}$}                    \\
 & & & & & & & & \\
\hline
 & & & & & & & & \\
CO             & 0$\rightarrow$1 & 1000 & $3.64 \pm 0.06$ & 2.8        
& 0.063 & 0.06 & Total &
$10^{+2}_{-2}$       & $2.3^{+0.8}_{-0.7} \times 10^{16}$ \\
               & 0$\rightarrow$1 & 240  & $3.07 \pm 0.11$ & 2.5        & 0.077 & 0.13 & Apr95(I) & & \\
               & 0$\rightarrow$1 & 495  & $3.62 \pm 0.12$ & $>$2.0     & 0.138 & 0.06 & May94(I) & & \\
               & 0$\rightarrow$1 & 338  & $3.78 \pm 0.08$ & $>$2.4     & 0.097 & 0.06 & Mar94(I) & & \\
               & 0$\rightarrow$1 & 230  & $3.08 \pm 0.36$ & $>$1.4     & 0.290 & 0.13 & Feb94(I) & & \\
 & & & & & & & & \\
$^{13}$CO      & 0$\rightarrow$1 & 1000 & $<$0.14         & $<$0.04    & 0.043 & 0.10 & Total &
$10^{+2}_{-2}$       & $<9.2^{+1.1}_{-2.8} \times 10^{14}$ \\
               & 0$\rightarrow$1 &  240 & $<$0.23         & $<$0.06    & 0.019 & 1.06 & Apr95(I) & & \\
               & 0$\rightarrow$1 & 512  & $<$0.19         & $<$0.07    & 0.022 & 0.72 & May94(I) & & \\
 & & & & & & & & \\
\hline
 & & & & & & & & \\
HCN            & 0$\rightarrow$1 &   75 & $<$0.92         & 0.06       & 0.063 & 2.08 & Feb96(K) & & \\
 & & & & & & & & \\
               & 1$\rightarrow$2 & 1000 & $0.99 \pm 0.07$ & 0.5        & 0.042 & 0.21 & Total &
$11.8^{+3.1}_{-2.1}$ & $7.3^{+2.7}_{-1.5} \times 10^{12}$ \\
               & 1$\rightarrow$2 &  170 & $1.52 \pm 0.14$ & 0.6        & 0.077 & 0.21 & Apr95(I) & & \\
               & 1$\rightarrow$2 &  320 & $1.36 \pm 0.11$ & 0.6        & 0.079 & 0.16 & May94(I) & & \\
               & 1$\rightarrow$2 &  243 & $0.46 \pm 0.10$ & 0.4        & 0.079 & 0.24 & Mar94(I) & & \\
 & & & & & & & & \\
               & 3--2 & 1000 & $0.93 \pm 0.10$ & 0.6        & 0.063 & 0.22 & Total &
$11.8^{+3.1}_{-2.1}$ & $7.3^{+1.1}_{-0.3} \times 10^{12}$ \\
               & 2$\rightarrow$3 &  120 & $0.95 \pm 0.53$ & 0.4        & 0.233 & 0.33 & Apr95(I) & & \\
               & 2$\rightarrow$3 &  173 & $1.53 \pm 0.19$ & 0.8        & 0.135 & 0.16 & May94(I) & & \\
               & 2$\rightarrow$3 &  146 & $0.80 \pm 0.10$ & 0.6        & 0.085 & 0.16 & Mar94(I) & & \\
 & & & & & & & & \\
\hline
 & & & & & & & & \\
HCO$^+$        & 0$\rightarrow$1 &   46 & $5.20 \pm 0.50$ & 1.5        & 0.218 & 0.41 & Feb96(K) &
$5.4^{+0.7}_{-0.5}$& \\
 & & & & & & & & \\
               & 1$\rightarrow$2 & 1000 & $6.87 \pm 0.06$ & $>$2.6     & 0.079 & 0.04 & Total &
$8.7^{+0.2}_{-0.2}$ & $2.9^{+0.1}_{-0.2} \times 10^{13}$ \\
               & 1$\rightarrow$2 &  170 & $5.57 \pm 0.19$ & $>$2.0     & 0.131 & 0.21 & Apr95(I) & & \\
               & 1$\rightarrow$2 &  340 & $7.22 \pm 0.11$ & $>$2.7     & 0.070 & 0.12 & May94(I) & & \\
               & 1$\rightarrow$2 &  242 & $6.20 \pm 0.15$ & $>$2.3     & 0.108 & 0.16 & Mar94(I) & & \\
               & 1$\rightarrow$2 &  128 & $4.51 \pm 0.32$ & $>$1.4     & 0.268 & 0.12 & Feb94(I) & & \\
 & & & & & & & & \\
               & 2$\rightarrow$3 & 1000 & $4.75 \pm 0.10$ & 2.0        & 0.122 & 0.06 & Total &
$8.7^{+0.2}_{-0.2}$ & $2.9^{+0.05}_{-0.05} \times 10^{13}$ \\
               & 2$\rightarrow$3 &  120 & $5.98 \pm 0.19$ & 2.3        & 0.085 & 0.32 & Apr95(I) & & \\
               & 2$\rightarrow$3 &  172 & $4.01 \pm 0.18$ & 1.3        & 0.172 & 0.11 & May94(I) & & \\
               & 2$\rightarrow$3 &  144 & $4.72 \pm 0.37$ & $>$1.6     & 0.230 & 0.22 & Mar94(I) & & \\
 & & & & & & & & \\
H$^{13}$CO$^+$ & 1$\rightarrow$2 &  329 & $<$0.18         & $<$0.08    & 0.027 & 0.42 & May94(I) &
$8.7^{+0.2}_{-0.2}$ & $<7.9^{+0.1}_{-0.2} \times 10^{11}$ \\
 & & & & & & & & \\
\hline
 & & & & & & & & \\
HNC            & 1$\rightarrow$2 & 1000 & $0.53 \pm 0.05$ & 0.3        & 0.040 & 0.21 & Total &
$15.5^{+9.5}_{-5.0}$ & $5.2^{+5.7}_{-2.0} \times 10^{12}$ \\
               & 1$\rightarrow$2 &  170 & $0.68 \pm 0.11$ & 0.3        & 0.043 & 0.41 & Apr95(I) & & \\
               & 1$\rightarrow$2 &  310 & $0.69 \pm 0.09$ & 0.3        & 0.078 & 0.12 & May94(I) & & \\
               & 1$\rightarrow$2 &  238 & $0.45 \pm 0.07$ & 0.2        & 0.054 & 0.16 & Mar94(I) & & \\
 & & & & & & & & \\
               & 2$\rightarrow$3 & 1000 & $0.60 \pm 0.11$ & 0.3        & 0.085 & 0.21 & Total &
$15.5^{+9.5}_{-5.0}$ & $5.2^{+3.2}_{-0.9} \times 10^{12}$ \\
               & 2$\rightarrow$3 &  120 & $0.93 \pm 0.29$ & ---        & 0.127 & 0.32 & Apr95(I) & & \\
               & 2$\rightarrow$3 &  168 & $0.78 \pm 0.18$ & 0.4        & 0.114 & 0.21 & May94(I) & & \\
 & & & & & & & & \\
\hline
 & & & & & & & & \\
CS             & 1$\rightarrow$2 &   80 & $<0.62$         & $<$0.21     & 0.048 & 1.88 & Feb96(K) &
$10^{+2}_{-2}$       & $<2.0^{+0.7}_{-0.6} \times 10^{13}$ \\
 & & & & & & & & \\
\hline
 & & & & & & & & \\
CN$^{b)}$ & 0$\rightarrow$1 & 1000 & $1.10\pm 0.23$ & 0.1& 0.012 & 3.29 &Total &
$10^{+2}_{-2}$       & $2.0^{+1.2}_{-1.1} \times 10^{13}$ \\
               & 0$\rightarrow$1 &  240 & $<0.19$         & $<$0.03     & 0.031 & 1.03 & Apr95(I) & & \\
               & 0$\rightarrow$1 &  341 & $<0.15$         & $<$0.01     & 0.014 & 3.29 & Mar94(I) & & \\
 & 0$\rightarrow$1 & 150 & $1.10 \pm 0.23$ & 0.1 & 0.012 & 3.29 & Jul96(I) &&\\
 & & & & & & & & \\
\hline
 & & & & & & & & \\
N$_2$H$^+$     & 0$\rightarrow$1 &  075 & $<1.07$         & $<$0.10     & 0.103 & 1.00 & Feb96(K) &
$10^{+2}_{-2}$       & $<9.9^{+3.7}_{-3.1} \times 10^{12}$ \\
 & & & & & & & & \\
               & 1$\rightarrow$2 &  170 & $<0.48$         & $<$0.03     & 0.028 & 2.00 & Apr95(I) &
$10^{+2}_{-2}$       & $<2.1^{+0.5}_{-0.4} \times 10^{12}$ \\
 & & & & & & & & \\
               & 2$\rightarrow$3 &  120 & $<1.1$          & $<$0.08     & 0.077 & 1.34 & Apr95(I) &
$10^{+2}_{-2}$       & $<6.3^{+0.3}_{+0.3} \times 10^{12}$ \\
 & & & & & & & & \\
\hline
 & & & & & & & & \\
H$_2$CO        & 1$_{1,1}\rightarrow 2_{1,2}$ &  296 & $<0.14$ & $<$0.03     & 0.031 &
0.55 & Mar94(I) & $10^{+2}_{-2}$       & $<1.5^{+1.3}_{-1.0} \times 10^{13}$ \\
 & & & & & & & & \\
\hline
 & & & & & & & & \\
O$_2$          & 1$_0$$\rightarrow$1$_1$ &  484 & $<0.14$ & $<$0.03     & 0.031 & 0.55 & May94(I) &
$10^{+2}_{-2}$       & $<2.0^{+0.7}_{-0.6} \times 10^{16}$ \\
 & & & & & & & & \\
\hline
\end{tabular}
\ \\
a)\ The continuum level in mK. Total averages have $T_{\rm c}=1000$\,mK. \\
b)\ The maximum optical depth. A lower limit is set when the absorption
reaches within the spectral noise rms of the zero level. \\
c)\ Channel--to--channel noise rms of the opacity. \\
d)\ Channel separation used for the derivation of opacities. \\
e)\ Derived or assumed excitation temperature in K. \\
f)\ Derived for an upper limit to the N$=1-0$, J$=3/2-1/2$, F$=5/2-3/2$ line.
\end{flushleft}
\end{table*}

\begin{figure}
\psfig{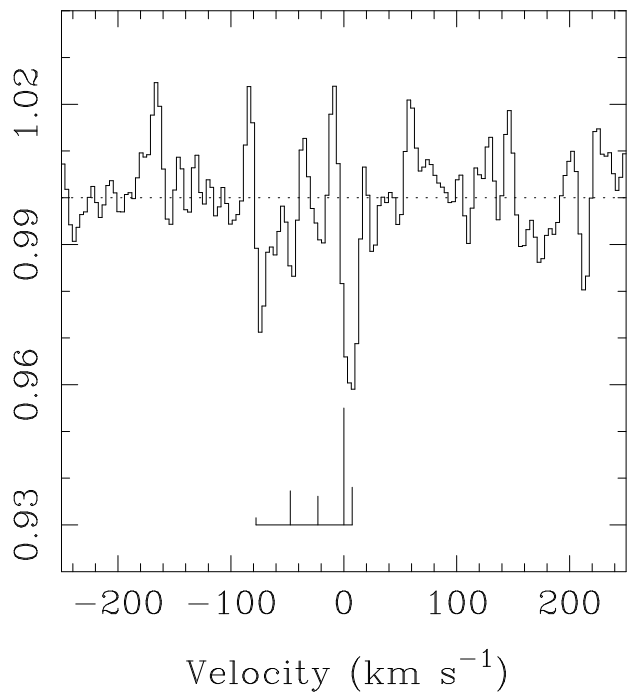}
\caption[]{Normalized spectra of CN N$=$0$\rightarrow$1 lines. From 
left to right we have indicated the J$=3/2-1/2$, F$=1/2-3/2$,
F$=3/2-3/2$, F$=1/2-1/2$, F$=5/2-3/2$ and F$=3/2-1/2$ hyperfine transitions.
Only the F$=5/2-3/2$ line is formally detected, possibly blended with the
F$=3/2-1/2$ line. The height of the markings corresponds to the observed
intensities in the Orion hot core (cf. Lovas 1992).
The velocity resolution is 3.3\,\kms\ and zero velocity corresponds to a
heliocentric redshift of 0.2467.}
\end{figure}

\begin{figure}
\psfig{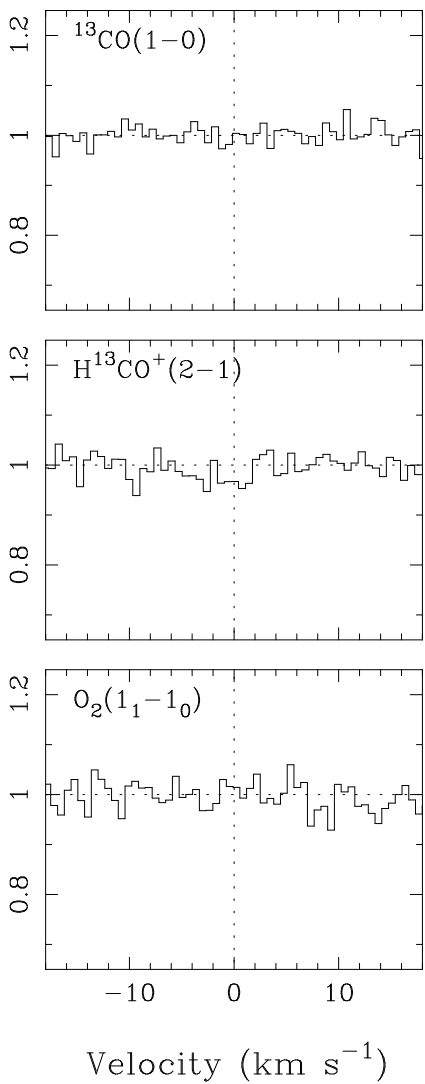}
\caption[]{Normalized spectra of the non--detected $^{13}$CO(1--0),
H$^{13}$CO$^+$(1$\rightarrow$2) and O$_2$(1$_{0}\rightarrow 1_{1}$)
lines. The velocity resolution is 0.65, 0.66 and 0.64\,\kms,
respectively. The velocity scale is heliocentric.
Notice the different scale for the continuum level compared with Fig.\,1.}
\end{figure}

\subsection{Derivation of physical parameters}

In order to derive physical parameters from molecular
absorption line data one must consider three uncertainties:
{\em (i)} the covering factor of absorbing gas across the
finite extent of the background continuum source,
{\em (ii)} the excitation temperature of the molecular gas
and {\em (iii)} if LTE conditions apply.
The first uncertainty affects the determination of the velocity
integrated optical depth while the second and the third also
affects the derivation of column densities.

\subsubsection{Opacity}

In a manner similar to Wiklind \& Combes (1994) we express
the observed continuum temperature, $T_{\rm c}$, away from
an absorption line as
\begin{equation}
T_{\rm c}=f_{\rm s}J(T_{\rm b}),
\end{equation}
where $f_{\rm s}$ is the beam filling factor of the region
emitting continuum radiation, $T_{\rm b}$ is the brightness
temperature of the background source and
$J(T)=(h\nu/k)/[1-\exp(-h\nu/kT)]$.
The spatial extent of the region emitting continuum radiation
at millimeter wavelengths is unknown. It is generally accepted
that as the wavelength decreases, the emission originates from
regions closer to the central AGN. At centimeter wavelengths,
the core of PKS1413+135 is unresolved at milliarcsecond (mas)
scales (Perlman et al. 1996). The core is likely to be smaller
at millimeter wavelengths. The BL Lac 3C446 has been observed
with mm--VLBI and has an extent $<30\,\mu$arcseconds (Lerner
et al. 1993).
Since the angular size of the telescope
beam is typically 25'', the brightness temperature of the
background source $T_{\rm b}$ is at least $10^{9} \times T_{\rm c}$.
This means that we do not have to consider the local excitation
temperature of the molecular gas when deriving the opacity,
since this is typically of the order 10\,K. The excitation
does enter, however, when deriving column densitites.

The observed antenna temperature of an absorption line can
be expressed as
\begin{equation}
\Delta T_{\rm A}^{*}(\nu)=f_{\rm s}J(T_{\rm b}) \{1-f_{\rm c} 
[1-exp(-\tau_{\nu})]\},    
\end{equation}
where $f_{\rm c}$ is the fraction of the continuum source
area covered by molecular gas at the velocity corresponding
to the frequency $\nu$. We here assume that $f_{\rm c}$ is
a characteristic covering factor for the molecular gas, independent
of the opacity $\tau_{\nu}$ of the gas.
$\Delta T_{\rm A}^{*}=T_{\rm c}-|T_{\rm abs}|$, where $T_{\rm abs}$
is the depth of the absorption line measured from the continuum
level. When $\tau_{\nu}$ is very large the absorption saturates, i.e.
$T_{\rm abs}^{*} \rightarrow T_{\rm c}$ and
$\Delta T_{\rm A}^{*} \rightarrow 0$.
In this equation, we cannot derive independently the covering factor
of the absorbing clouds, $f_{\rm c}$, and their optical depth.
Combining Eqs (1) and (2) we can express the `true' optical
depth as:
\begin{eqnarray}
\tau_{\nu} & = & - \ln [1-\frac{1}{f_{\rm c}}
(1-\exp(-\tau_{\nu_{\rm obs}}))]\ , \nonumber \\
{\rm where} \\ 
\tau_{\nu_{\rm obs}} & = & \ln (\frac{T_{\rm c}}{\Delta T_{\rm A}^{*}})\ . \nonumber
\end{eqnarray}
$\tau_{\nu_{\rm obs}}$ is the optical depth averaged over
the size of the background continuum source and over the
excitation conditions of the molecular gas along the line
of sight. This value is given for each observed absorption
line in Table\,2. By assuming that the opacity of the
absorbing gas is high, Eq. (3) gives a minimum possible
value for the covering factor
$f_{\rm c} \ga 1 - \exp(-\tau_{\nu_{\rm obs}})$. Should the
opacity be low, the filling factor must be even larger.
The velocity integrated opacity is obtained by integrating
$\tau_{\nu}$ in Eq.(4) over the full extent of the
absorption line.

\medskip

In PKS1413+135 the optical depth of the CO(0$\rightarrow$1) and
HCO$^+$(1$\rightarrow$2) lines is $\sim$3. This means that
the absorption almost reaches the zero level, but it is
not heavily saturated. The HCN, HNC and the J$=$2$\rightarrow$3
line of HCO$^+$ all have smaller optical depths.
The non--detections of less abundant molecules such as CS,
N$_2$H$^+$ and the isotopic variants $^{13}$CO and H$^{13}$CO$^+$
are consistent with the detected lines not being heavily saturated.
The large depth of the CO(0$\rightarrow$1) and
HCO$^+$(1$\rightarrow$2) lines then implies that the covering factor
$f_{\rm c}$ must be very close to unity. In the following we
will set $f_{\rm c}=1$ for PKS1413+135.

\begin{figure}
\psfig{figure=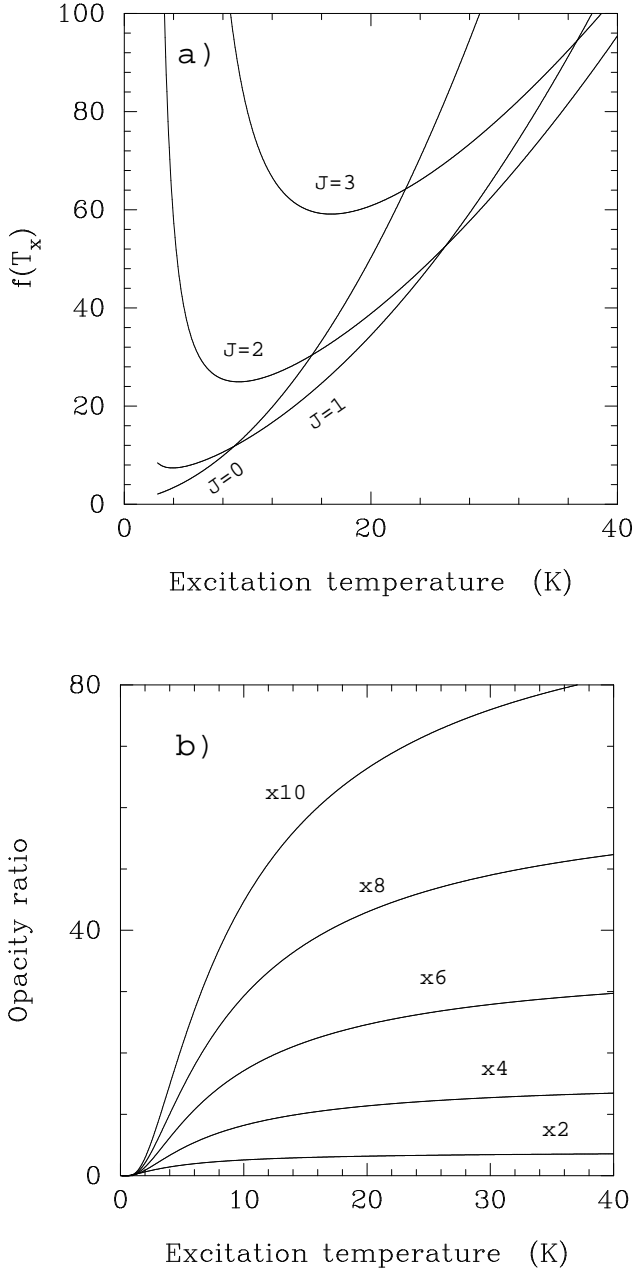,bbllx=60mm,bblly=60mm,bburx=143mm,bbury=235mm,width=8.5cm,angle=0}
\caption[]{{\bf a)}\ The function $f(T_{\rm x})$ for the 4 lowest rotational transitions
of HCO$^+$ as a function of the excitation temperature $T_{\rm x}$ (cf. Eq.\,6).
{\bf b)}\ The ratio of integrated optical depth for the J$=$1$\rightarrow$2 line of
HCO$^+$ for two different excitation temperatures, $T_1$ and $T_2$. The number at each
line in the plot corresponds to the factor $a$, $T_2 = aT_1$. The results shows that
for a given transition and equal column densities, the coldest gas component contributes
the largest part of the observed opacity.}
\end{figure}

\begin{figure}
\psfig{figure=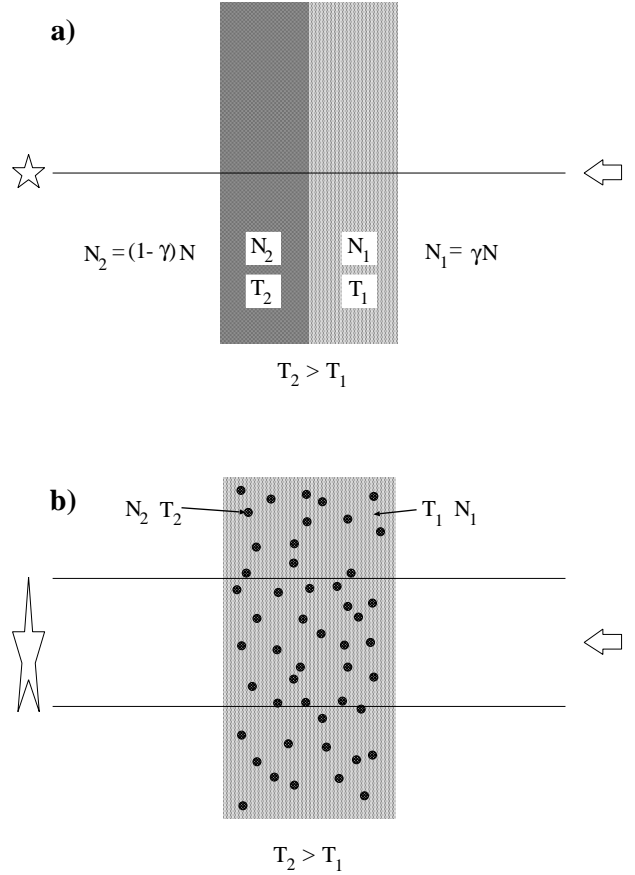,bbllx=10mm,bblly=10mm,bburx=200mm,bbury=280mm,width=8.5cm,angle=0}
\caption[]{{\bf a)} A schematic illustration of a two--component molecular medium seen in
absorption. The two components are characterized by excitation temperatures
$T_1$ and $T_2$, with $T_2 > T_1$ and represents a fraction $\gamma$ and
$1-\gamma$ of the total column density $N$ (see the text for more details); 
{\bf b)} Same as a) but with the warmer (denser) gas distributed in
a clumpy medium. The line of sight is shown `extended' to emphasize the
clumpiness present within the gas causing the absorption profile.}
\end{figure}

\begin{figure}
\psfig{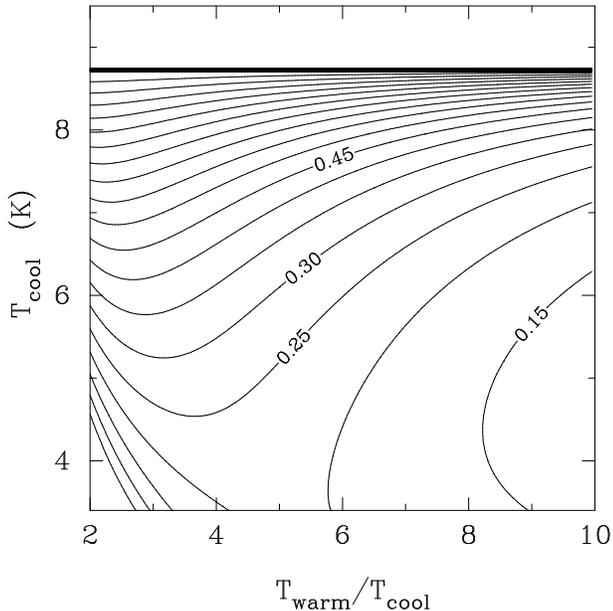}
\caption[]{The fraction $\gamma$ of cool molecular gas in a two--component 
model, covering both all the continuum surface,
of the molecular gas along the line of sight to PKS1413+135. 
The fraction is derived from the ratio of the J$=$1$\rightarrow$2
and J$=$2$\rightarrow$3 transitions of HCO$^+$.}
\end{figure}

\begin{figure}
\psfig{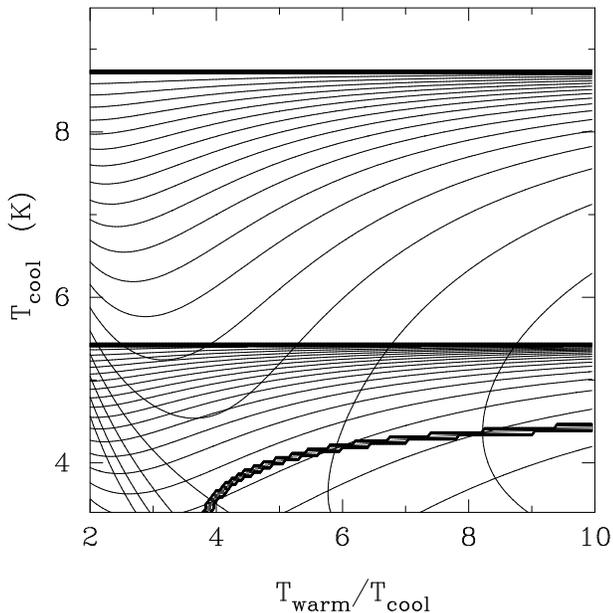}
\caption[]{The same fraction $\gamma$ as in Fig.\,6, but now including
solutions also for the J$=$0$\rightarrow$1/J$=$1$\rightarrow$2 ratio.
The grey--scaled area in the lower part of the figure represents the
region where the two solutions agree, suggesting $\gamma = 0.2$,
$T_{\rm cool}=4.2$\,K and $T_{\rm warm}\approx 25-30$\,K.}
\end{figure}

\subsubsection{Excitation temperature and column density}

The excitation temperature $T_{\rm x}$ relates the relative population
of two rotational levels of a molecule as:
\begin{equation}
\frac{n_{2}}{n_{1}} = \frac{g_{2}}{g_{1}}\ \exp\left(-h\nu_{21}/kT_{\rm x}\right),
\end{equation}
where $g_{\rm i}$ is the statistical weight for level $i$ and
$h\nu_{21}$ is the energy difference between two rotational levels.
In order to derive $T_{\rm x}$ we must link the fractional population
in level $i$ to the total abundance (here abundance is used in synonym
with column density). This is done by invoking a weak LTE--approximation
and assuming that $T_{\rm x} = T_{\rm rot}$, where $T_{\rm rot}$ is
a temperature which governs the fractional population of all rotational
levels in a given molecule. The LTE approximation is weak in the sense
that it does not imply that $T_{\rm rot}$ equals the kinetic temperature
and it allows for different molecules to have different $T_{\rm rot}$. 
With this approximation we can use the partition function
$Q(T_{\rm x})=\sum_{J=0}^{\infty}g_{\rm J}\exp(-E_{\rm J}/kT_{\rm x})$
to express the total column density $N_{\rm tot}$ as
\begin{eqnarray}
N_{\rm tot} & = & {8\pi \over c^{3}} {\nu^{3} \over g_{J} A_{J,J+1}} f(T_{\rm x})
             \ \int \tau_{\nu} dV\ ,\nonumber \\
\ \\
f(T_{\rm x}) & = & {Q(T_{\rm x}) \exp(E_{J}/kT_{\rm x}) \over
1-\exp(-h\nu/kT_{\rm x})}\ ,\nonumber
\end{eqnarray}
where $\int \tau_{\nu} dV$ is the observed optical depth integrated
over the line for a given transition, $g_{\rm J}=2J+3$ for a
transition $J \rightarrow J+1$ and $E_{\rm J}$ is the energy of
the rotational level $J$.
By taking the ratio of two observed transitions from the same molecule, the
excitation temperature can be derived.
If the lower transition is saturated, $T_{\rm x}$ will be an upper limit.
Non--saturated lines will directly give $T_{\rm x}$, limited only by the S/N
in the spectra.
The derivation of the excitation temperature depends only  on the
line--to--continuum ratio of the two transitions in question and is
therefore independent of a correctly measured continuum level, as
long is this is not caused by changing weather conditions inbetween
two calibrations.
The $T_{\rm x}$ derived in this way is an average
excitation temperature along the line of sight and averaged over the
finite extent of the continuum source.
The strong frequency dependence of the column density in Eq,(6) is
only apparent since the Einstein coefficient $A_{\rm J,J+1}$ is
proportional to $\nu^3$.

\subsection{Multicomponent analysis}

Towards PKS1413+135 we have observed three transitions of HCO$+$ which
allows us to derive the excitation temperature using the ratio of the
observed integrated opacities for the
J$=$0$\rightarrow$1/J$=$1$\rightarrow$2 and
J$=$1$\rightarrow$2/J$=$2$\rightarrow$3
lines. If weak--LTE conditions apply, as defined above, these ratios
should give the same excitation temperature. This, however, is not
the case. The first ratio gives $T_{\rm x}=5.4^{+0.7}_{-0.5}$\,K,
the second $8.7^{+0.2}_{-0.2}$\,K.
Is this an indicaton that the assumption of LTE conditions are
at fault?

The above derivation of the excitation temperature and column
density corresponds to values averaged over the angular extent
of the background source as well as along the line of sight.
Since these extents are both finite, we may well have several
gas components characterized by different excitation temperatures
contributing to the observed absorption lines.

The observed property of a molecular absorption line is the
velocity integrated optical depth $\int \tau_{\nu} dV$,
which is proportional to $N_{\rm tot}/f(T_{\rm x})$ (e.g. Eq.\,6).
As shown in Fig.\,4a, the function $f(T_{\rm x})$ is highly
nonlinear. The ratio of the integrated opacities for a given
molecule and transition will therefore depend strongly on the
excitation temperature. The ratio of opacities of two gas
components characterized by excitation temperatures $T_1 < T_2$ is
shown in Fig.\,4b for
HCO$^+$(1$\rightarrow$2), for $T_2/T_1$ ratios of 2--10.
It is evident from the figure that the `cool' gas component
contributes significantly more to the opacity than the
`warm' gas. Note that the temperature refers to the excitation
conditions, an excitationally cold gas can still be kinetically
warm if the density is low enough. It is thus possible to combine
a relatively large column of warm (i.e. dense) molecular gas with
smaller column of excitationally cold (i.e. diffuse) gas to
produce the observed opacity.
In order to see how such a two--component medium can
combine and still be consistent with the observed data,
we express the opacity as:
\begin{eqnarray}
\int \tau_{\nu} dV = {c^3 \over 8\pi} {A_{\rm J,J+1} g_{\rm J+1} \over \nu^3}
N_{\rm tot} \left[{\gamma \over f(T_1)} + {1-\gamma \over f(T_2)}\right]\ ,
\end{eqnarray}
where $\gamma$ is the fraction of the total column density
characterized by a temperature $T_1$ and $1-\gamma$ is the
fraction associated with gas of a temperature $T_2$. In the
following we will assume that $T_1 < T_2$ and that the
covering factor of the two components are equal. A schematic
view of this two--component model is shown in Fig.\,5a.
In the next section we will study the case when the
assumption of equal filling factors is relaxed.
Taking the ratio of two observed opacities the column
density is eliminated and we can solve for $\gamma$ as
a function of $T_1$ and $T_2$. One result obtained for
the J$=$1$\rightarrow$2 and J$=$2$\rightarrow$3 lines of
HCO$^+$ is shown in Fig.\,6. The ratio of the observed
opacities give an average excitation temperature of 8.7\,K
(see Table\,2), which is the limit for $\gamma =1$ in
Fig.\,6. It is, however, possible to have a fraction
$\gamma<$1 of the total column density in the form of
significantly colder gas, with the remaining $1-\gamma$
in the form of a warmer gas component. The $x-$axis in
Fig.\,6 shows the ratio of the two temperatures.
Hence, one solution is to have 25\% of the column density
in the form of a gas component with $T_{\rm x}\approx$5\,K
and 75\% in the form of a warm gas component with
$T_{\rm x}\approx$20\,K.

By including all three transitions observed for HCO$^+$,
we get two sets of solutions for the fraction $\gamma$.
These are shown in Fig.\,7. The grey--scaled area
indicates where the two solutions coincide.
The results suggests the presence of an excitationally
cold component with $T_{\rm x} \approx 4$\,K which
constitutes a fraction $\sim$20\% of the total column
density. The remaining 80\% is made up of a gas with
a $T_{\rm x} \approx 20-40$\,K.
Hence, the non--concordant values of the excitation
temperatures derived from three transitions of HCO$^+$
do not necessarily imply that the assumption of weak--LTE
is violated, but are consistent with the presence of two
gas components with different excitation temperatures.

There might be more than two gas components along the line of
sight. By making the ansatz that we have N components, each
charaterized by an excitation temperature $T_{\rm i}$, with
$T_{1} < T_{2} < ... < T_{\rm N}$, the fraction $\gamma$ of
the coldest gas component will be somewhat smaller than derived
assuming a 2--component model. However, the strong nonlinear
behaviour of the opacity as a function of $T_{\rm x}$ (Fig.\,4a,b),
ensures that a third, fourth or higher gas component will give
a small or neglible contribution to the observed opacity.
{\em Thus, a 2--component model should give a reasonable
approximation to the possible gas components existing
along the line of sight and over the finite extent of the
background source.}

Adopting values for $T_1$ and $T_2$, giving a fraction $\gamma$,
the total column density $N$ can be derived. This value is always
larger than for the `average' $T_{\rm x}$, with a maximum for
the smallest allowed fraction $\gamma$ of cold gas.
This is a natural consequence of the nonlinear dependence of
the opacity to the excitation temperature as shown in Fig.\,4a.
In the case of HCO$^+$, the allowed solutions give total column
densities in the range $3.7-11.2 \times 10^{13}$\,cm$^{-2}$,
with corresponding excitation temperatures 3.4 \& 13.3\,K
and 4.5 \& 44.5\,K for the cold and warm components. The
fraction $\gamma$ is 0.19 and 0.12 for these two cases.
The column density derived using the average $T_{\rm x}$ as
given in Table\,2 is $2.9 \times 10^{13}$\,cm$^{-2}$.
The maximum allowed fraction of cold gas is $\gamma=0.22$,
with excitation temperatures 3.9 \& 18.0\,K and a total column
density of HCO$^+$ of $4.5 \times 10^{13}$\,cm$^{-2}$.

\begin{figure*}
\psfig{figure=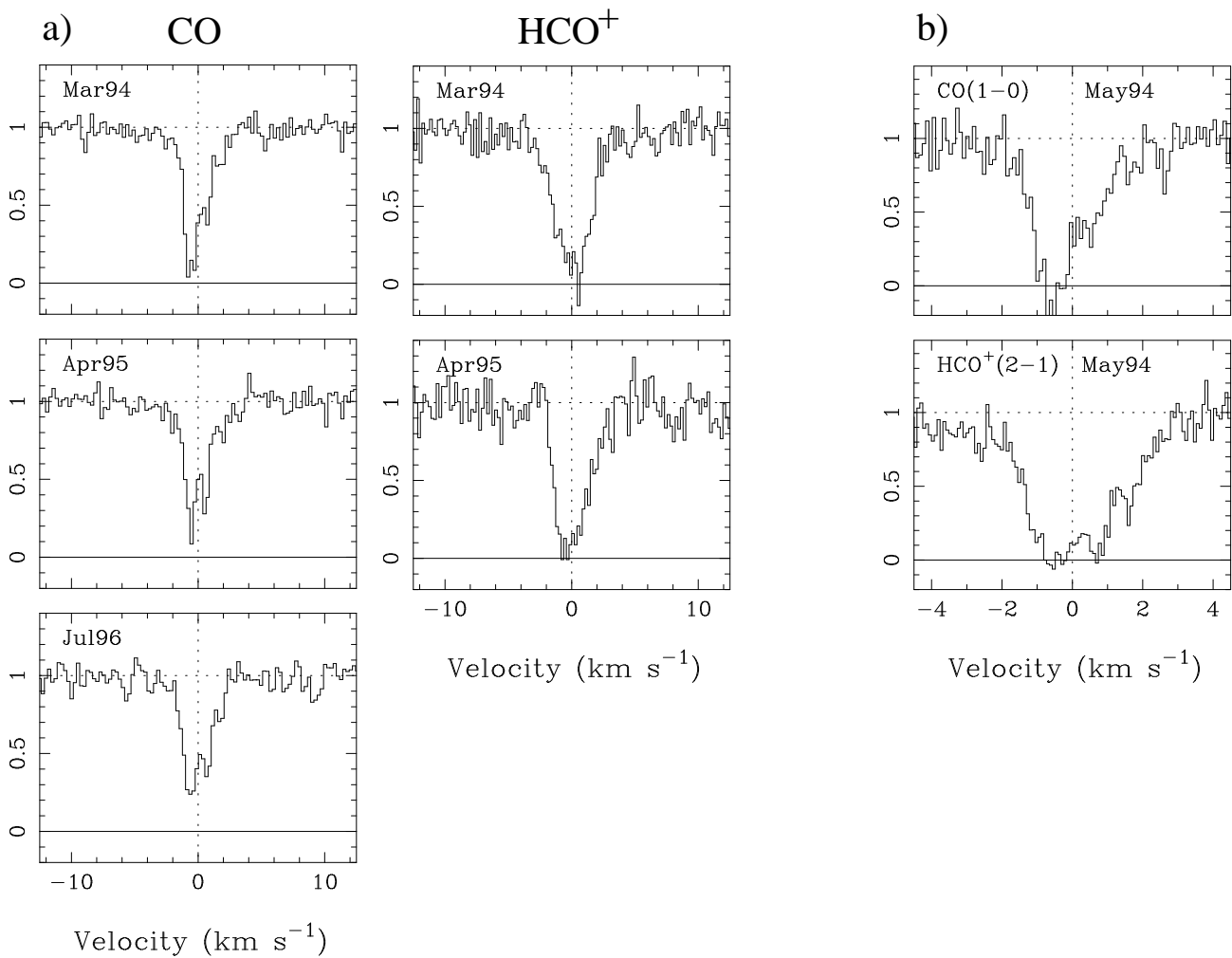,bbllx=8mm,bblly=65mm,bburx=198mm,bbury=215mm,width=17.5cm,angle=0}
\caption[]{{\bf a)}\ Normalized spectra of the J$=$0$\rightarrow$1
transition of CO obtained in March 1994, April 1995 and July 1996
and the J$=$1$\rightarrow$2 transition of HCO$^+$ obtained in
March 1994 and April 1995. The velocity resolution is 0.25\,\kms\ for
the CO spectra and 0.20\,\kms\ for the HCO$^+$ spectra.
{\bf b)}\ Normalized spectra of the J$=$0$\rightarrow$1 transition
of CO and the J$=$1$\rightarrow$2 transition of HCO$^+$ both
obtained in May 1994 with a velocity resolution of 0.095\,\kms\
and 0.082\,\kms, respectively.
The velocity scale is heliocentric and the dashed line corresponds
to zero velocity for a redshift $z=0.24671$.}
\end{figure*}

\subsection{A clumpy medium}

Since the excitation of molecular gas is generally governed
by collisions with H$_2$, a high-$T_x$ gas is characterized by a high 
density.The division in cool and warm gas made in the previous
section is therefore equivalent to a division into diffuse
and dense gas. A more realistic multi-phase medium combining 
the dense and diffuse phases is that observed in the Milky Way ISM,
where a clumpy structure of small and dense clouds with a small filling factor
is embedded in a pervasive diffuse molecular medium.
A schematic view of this two--component model is shown in
Fig.\,5b.

If we call $f_{\rm c}$ the area filling factor of each clump, for which
$\tau_{\nu} >> 1$, their contribution to the absorption depth is simply 
$f_{\rm c}$ (assuming that the brightness temperature of the
background source is homogeneous over its extent).
If we assume that there are in total $n_{\rm c}$ clouds in the beam,
and that their line-width is  $\delta v$ each,
corresponding to a velocity filling factor $f_v= \delta v/\Delta V$
(where $\Delta V$ is the total spectrum line-width observed), and
neglecting cloud overlapping for the sake of simplicity, 
their combined contribution to the opacity, at at given $\nu$, will be
\begin{equation}
n_{\rm c}f_v f_{\rm c} = 1 - \exp(-\tau_{\rm obs})\ ,
\end{equation}
where $\tau_{\rm obs}$ is the observed opacity.
The total H$_2$ column density in the beam is then
\begin{equation}
N_{\rm H_2} = \frac{4}{3} n_{\rm H_2} r_{\rm c}
{ 1 - \exp(-\tau_{\rm obs}) \over f_{\rm c}f_v}
\ \ {\rm cm}^{-2} \ ,
\end{equation}
where $n_{\rm H_2}$ is the density of molecular hydrogen
in each clump and $r_{\rm c}$ the characteristic radius
of the clumps.
In order to see how the presence of small dense structures
influence the derivation of column density we can use the
simple parametrization presented above, together with the
characteristic values derived from observations of the
Galactic molecular ISM.
Assuming a $\tau_{\rm obs} = 0.5 $ for the
CO(0$\rightarrow$1) line a velocity width
$\delta v = 0.1$\,km\,s$^{-1}$ and a covering factor
$f_{\rm c} = 0.001$ per clump, the number of small clumps
in the beam will be
$n_{\rm c} \approx 4 \times 10^3$. If we further assume a radius of
100\,AU for the clumps (approximately corresponding
to 0.1\% of the area of a background source with a diameter
of 6000\,AU) and a density $n_{\rm H_2} = 10^{4}$\,cm$^{-3}$,
Eq.\,(10) gives a column density
$N(H_2) \approx 8 \times 10^{22}$\,cm$^{-2}$. The
corresponding CO column density, derived solely from the
observed opacity and assuming $f_{\rm c} \approx 1$,
would in this case be
$3 \times 10^{15}$\,cm$^{-2}$, which translates to a
H$_2$ column density of $6 \times 10^{19}$\,cm$^{-2}$.
This simple excercise shows that {\em the column density can
be underestimated by three orders of magnitude if the
molecular ISM is made up of small dense clumps rather
than a diffuse medium}.

\begin{table}
\begin{flushleft}
\caption[]{Column densities}
\small
\begin{tabular}{ccccc}
\hline
 & & & & \\
\multicolumn{1}{c}{Molecule}                     &
\multicolumn{1}{c}{$\gamma^{a)}$}                &
\multicolumn{1}{c}{$T_{\rm cold, diffuse}$}      &
\multicolumn{1}{c}{$T_{\rm warm, dense}$}        &
\multicolumn{1}{c}{$N$}                          \\
 & & & & \\
\multicolumn{1}{c}{ }                            &
\multicolumn{1}{c}{ }                            &
\multicolumn{1}{c}{K}                            &
\multicolumn{1}{c}{K}                            &
\multicolumn{1}{c}{cm$^{-2}$}                    \\
 & & & & \\
\hline
 & & & & \\
CO      & 0.22 & 3.9 & 18.0 & $2.0 \times 10^{16}$ \\
 & & & & \\
HCN     & 0.05 & 3.4 & 27.0 & $2.0 \times 10^{13}$ \\
 & & & & \\
HNC     & 0.05 & 3.4 & 27.0 & $1.0 \times 10^{13}$ \\
 & & & & \\
HCO$^+$ & 0.22 & 3.9 & 18.0 & $4.5 \times 10^{13}$ \\
 & & & & \\
\hline
\end{tabular}
\ \\
a)\ The fraction of the column density in the cold gas component.
\end{flushleft}
\end{table}

\begin{table*}
\begin{flushleft}
\caption[]{Gaussfits to the CO(0$\rightarrow$1) line.}
\scriptsize
\begin{tabular}{cccccccccc}
\hline
 & & & & & & & & & \\
\multicolumn{1}{c}{Date}                                 &
\multicolumn{1}{c}{$T_{\rm c}^{a)}$}                     &
\multicolumn{1}{c}{Comp.}                                &
\multicolumn{1}{c}{$T_{\rm abs}^{b)}$}                   &
\multicolumn{1}{c}{$V_{\rm c}^{c)}$}                     &
\multicolumn{1}{c}{$\Delta V_{1/2}^{d)}$}                &
\multicolumn{1}{c}{$\chi^2$}                             &
\multicolumn{1}{c}{$\left[\int{\tau_{\nu} dV}\right]_{i}^{e)}$}       &
\multicolumn{1}{c}{$\left[\int{\tau_{\nu} dV}\right]_{\rm tot}^{f)}$} &
\multicolumn{1}{c}{$R^{g)}$}                             \\
 & & & & & & & & & \\
\multicolumn{1}{c}{}                                     &
\multicolumn{1}{c}{}                                     &
\multicolumn{1}{c}{}                                     &
\multicolumn{1}{c}{}                                     &
\multicolumn{1}{c}{km\,s$^{-1}$}                         &
\multicolumn{1}{c}{km\,s$^{-1}$}                         &
\multicolumn{1}{c}{}                                     &
\multicolumn{1}{c}{km\,s$^{-1}$}                         &
\multicolumn{1}{c}{km\,s$^{-1}$}                         &
\multicolumn{1}{c}{}                                     \\
 & & & & & & & & & \\
\hline
 & & & & & & & & & \\
Feb94 & 0.46 & 1 & 0.91 & $-$0.7 & 0.9 & 4.87 & 2.6 & $3.2 \pm 0.2$   &
--- \\
      &      & 2 & 0.38 & +0.5   & 1.9 &      & 1.5 &                 &
    \\
      &      & 3 & 0.07 & +1.1   & 7.4 &      & 0.9 &                 &
    \\
 & & & & & & & & & \\
Mar94 & 0.68 & 1 & 0.97 & $-$0.5 & 1.3 & 0.44 & 2.8  & $3.80 \pm 0.05$ &
$5.6^{+0.8}_{-0.8}$ \\
      &      & 2 & 0.56 &   +0.7 & 0.6 &      & 0.48 &                 &
    \\
      &      & 3 & 0.25 &   +1.8 & 1.1 &      & 0.31 &                 &
    \\
 & & & & & & & & & \\
May94 & 1.00 & 1 & 1.10 & $-$0.6 & 1.2 & 0.83 & $>$3.3  & $4.07 \pm 0.07$ &
$>7.0^{+0}_{-0.8}$  \\
      &      & 2 & 0.47 &   +0.6 & 0.8 &      & 0.47    &              &
    \\
      &      & 3 & 0.15 &   +1.6 & 1.9 &      & 0.32    &              &
    \\
 & & & & & & & & & \\
Apr95 & 0.48 & 1 & 0.83 & $-$0.6 & 1.1 & 0.31 & 1.6  & $3.00 \pm 0.06$ &
$3.3^{+0.3}_{-0.1}$ \\
      &      & 2 & 0.62 &   +0.6 & 0.6 &      & 0.49 &                 &
    \\
      &      & 3 & 0.20 &   +1.7 & 1.9 &      & 0.42 &                 &
    \\
 & & & & & & & & & \\
Jul96 & 0.31 & 1 & 0.77 & $-$0.6 & 1.5 & 0.60 & 1.9 & $2.98 \pm 0.06$ &
$3.1^{+0.2}_{-0.1}$ \\
      &      & 2 & 0.56 &   +0.8 & 0.8 &      & 0.61 &                 &
    \\
      &      & 3 & 0.31 &   +1.8 & 0.6 &      & 0.63 &                 &
    \\
 & & & & & & & & & \\
\hline
\end{tabular}
\ \\
a)\ Continuum flux normalized to the flux at May 1994. \\
b)\ The depth of the absorption line from the normalized continuum level. \\
c)\ The center velocity of the Gauss component. \\
d)\ Full width at half intensity. \\
e)\ Integrated opacity of individual components derived from the Gaussian fits. \\
f)\ Total integrated opacities from the observed profiles. \\
g)\ Ratio of integrated opacities of component 1 and 2. Errors correspond to
$\pm$10\% variation of the continuum level.
\end{flushleft}
\end{table*}

\begin{figure}
\psfig{figure=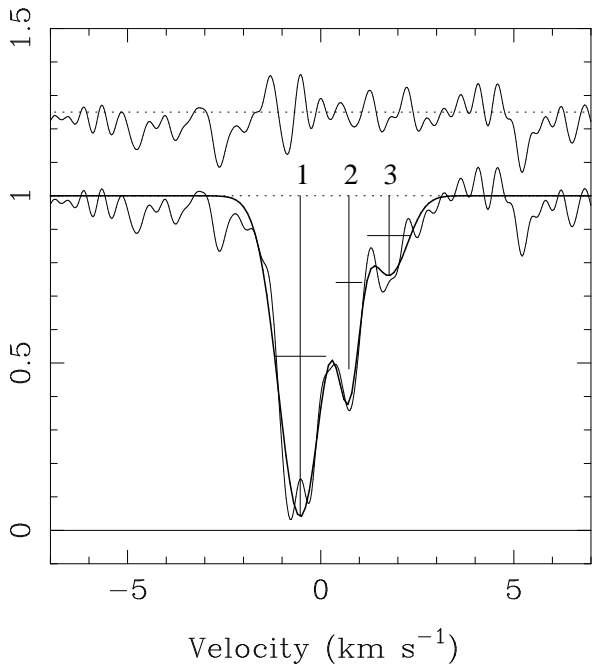,bbllx=55mm,bblly=90mm,bburx=155mm,bbury=200mm,width=8.5cm,angle=0}
\caption[]{Three component Gaussian fitted to the CO(0$\rightarrow$1)
absorption for the spectrum obtained in March 1994. The residual is 
shown above the spectrum. The parameters of the Gauss fit is given
in Table\,3. The numbers correspond to the components as listed in
Table\,3.}
\end{figure}

\subsection{Time variability}

In Fig.\,8 we show the CO(0$\rightarrow$1) and
HCO$^+$(1$\rightarrow$2) absorption lines for different
epochs, spanning 2 years for CO and 1 year for HCO$^+$.
In Fig.\,8b we show the CO and HCO$^+$ line at higher
velocity resolution.
While there are no apparent changes within the noise in the
HCO$^+$ spectra between March 1994 and April 1995, the CO
line appears to have changed. A closer inspection of the
CO absorption reveals that it is the deepest absorption
component that has changed while the second deepest component
remains more or less constant. In order to quantify possible
variations we fit 3 Gaussian components to the CO
spectra from all observed epochs. The components are
illustrated in Fig.\,9 and the Gaussian parameters are listed
in Table\,4, together with the integrated opacities derived
from the Gaussian fits. The data from February 1994 are of
poor quality, as seen from the high $\chi^2$ value of the
Gaussian fit, and will not be used in the analysis.
The integrated opacities show that while the second deepest
absorption component (no. 2 in Table\,4) does not show any
significant variation, the main component (no. 1)
changes by a factor of $>$2 in velocity integrated opacity.
The third component is much weaker and less clearly defined.
The Gaussian fit for this component is therefore more questionable
than for components 1 \& 2, which show consistent center
velocities and half widths for the different epochs.

How certain are these results? Changes in the integrated
opacity could be mimicked by an incorrectly derived
continuum level, if this is caused by calibration errors.
However, pure pointing uncertainties would not affect the
{\em ratio} of the components.
It is difficult to quantify the errors associated with
multicomponent Gaussian fits to absorption spectra.
The errors associated with the total velocity
integrated opacity are small (see Table\,4).
Hence, as long as the Gaussian fit is good, the error in the
integrated opacity for the separate components is small
($\la$5\%). The dominating uncertainty is associated with
errors in the continuum level caused by an unstable atmosphere.
This can be minimized by taking the ratio of integrated
opacities for the first and second
component (last column in Table\,4), but due to the
non--linear relation between the depth of the absorption
profile and the opacity a small uncertainty remains.
The continuum level is derived using $\ga$10 individual
spectra plus a number of special continuum observations.
Each of these has a typical uncertainty of $\pm$15\%.
The final uncertainty associated with the continuum level
from the average of these individual measurement
is approximately $\pm$5\%.
In order to quantify the uncertainty of the integrated
opacities introduced by errors in the continuum level,
we have done the analysis (Gaussian fits and derivation
of opacities) with a continuum level varying $\pm$10\%
around the observed value.
This is a conservative estimate and nevertheless shows
that the ratio of the integrated opacity of the first and
second absorption components has varied from $\ga$7 in May
1994 to $\sim$3 in 1995 and 1996. The ratio of $R$ between
May 1994 and July 1996, including the assumed errors of
$\pm$10\% in the continuum level, is $2.3 \pm 0.3$.

The fact that we do not see any clear changes in the HCO$^+$ line
is not contradictory. The opacity of the HCO$^+$
absorption is higher than that of CO, making it less
sensitive to changes in the continuum level. Moreover, the
abundance ratio HCO$^+$/CO is higher in low density environments
than at high densities (cf. Lucas \& Liszt 1996), with
the HCO$^+$ molecule existing in low density regions where
CO self--shielding is not taking place.

\begin{figure*}
\psfig{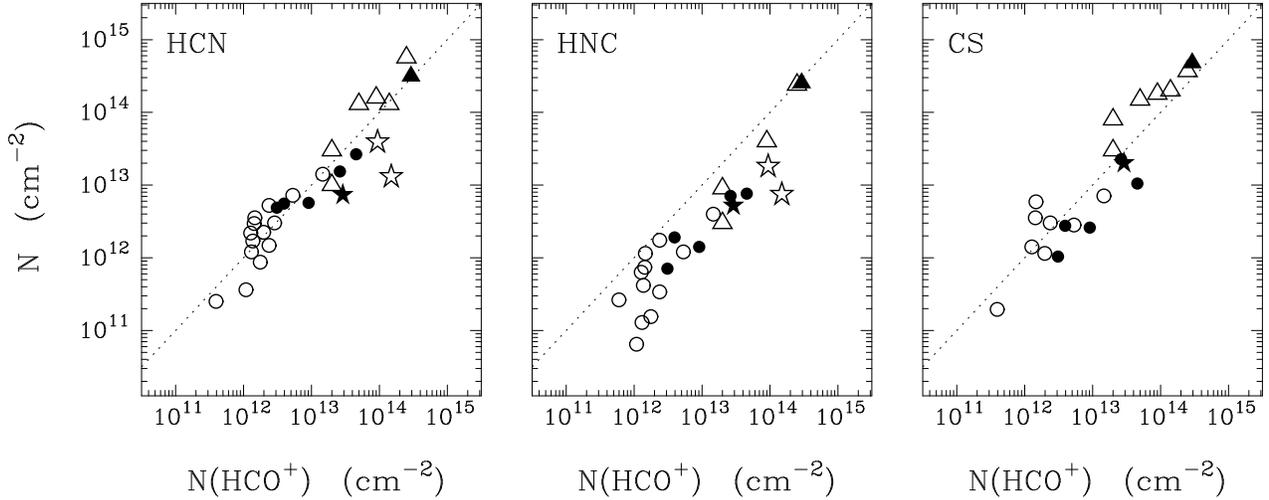}
\caption[]{Column density of HCO$^+$ plotted versus column densitites of HCN,
HNC and CS. The open circles represent Galactic diffuse clouds (Lucas \& Liszt
1994, 1996), filled circles represent data from Cen A (Wiklind \& Combes 1997)
and open triangles represent absorption data towards SgrB2 (Greaves \& Nyman
1996). The filled star represents our absorption data for PKS1413+135, open
stars from B3\,1504+377 at z$=$0.673 (Wiklind \& Combes 1996b) and the filled
triangle PKS1830--210 at z$=$0.886 (Wiklind \& Combes 1996a). The dashed line
is a one--to--one correspondance between the column densities and is not a
fit to the data.}
\end{figure*}

\begin{figure*}
\psfig{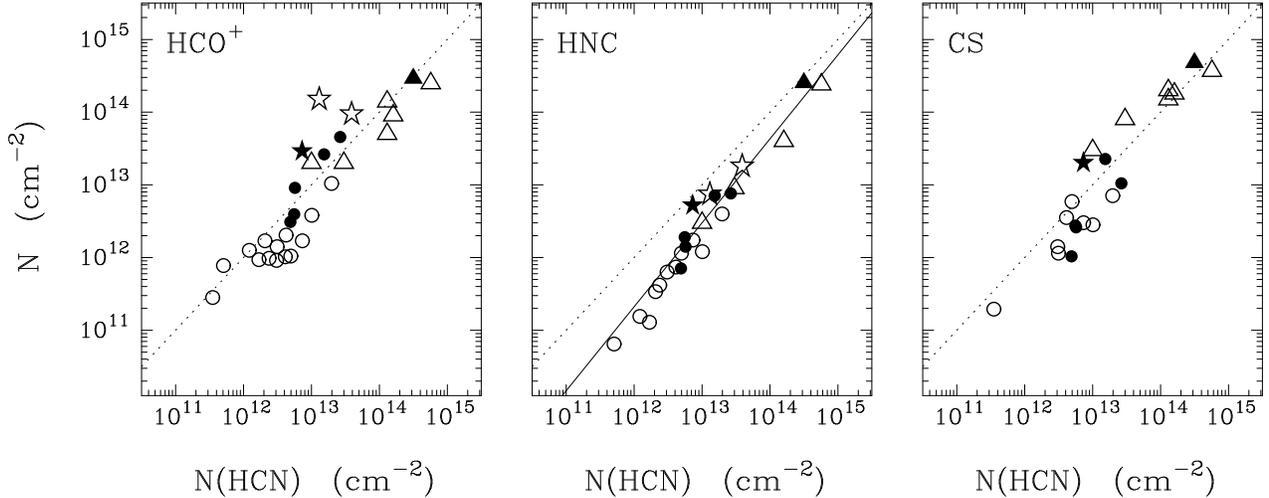}
\caption[]{Column density of HCN plotted versus column densitites of HCO$^+$,
HNC and CS. Designations are as in Fig.\,10. The full--drawn line in $N(HCN)$ vs.
$N(HNC)$ is a linear fit to the data, showing how the $HCN/HNC$ ratio increases
with decreasing HCN column density (see text for details).}
\end{figure*}

\section{Discussion}

\subsection{Small scale structure in the molecular gas}

In the Milky Way, several lines of evidence suggest that
the molecular ISM is clumpy on all scales from 20\,AU to
100\,pc and characterized by a fractal structure (cf.
Falgarone et al. 1992).
Time variations in Galactic H$_2$CO absorption towards
extragalactic point sources (Marscher et al 1993) even
suggest structures in the molecular ISM on scales of 10\,AU.
Through numerical simulations Marscher \& Stone (1994) were
able to derive constraints on the fractal structure of
molecular clouds from the time variability detections.
The mean number of small clumps along the line of sight
should be larger than previously thought, i.e. the size
spectrum of clumps should be a steeper power--law,
constraining the fractal dimension.
Detailed multilevel studies of CO and CS near the edge of
a nearby molecular cloud complex have revealed that the 
molecular ISM is confined to small structures (down to
35\,AU) which are dense ($n_{\rm H_2} \sim 10^{4}$\,cm$^{-2}$),
cold ($T_{\rm K} \sim 10-15$\,K) (Falgarone \& Phillips
1996) and characterized by a fractal structure (Falgarone et al.
1992). In fact, Falgarone \& Phillips (1996) find no evidence
for a pervasive diffuse molecular ISM, although this could be
a peculiarity for the specific region studied.
Although emission studies are not sensitive to the pervasive
diffuse medium, the absorption studies of Lucas \& Liszt (1996)
and Liszt \& Lucas (1996) show a higher rate of molecular
absorption in the Galaxy than was originally expected, and
precisely detect the low--opacity, large--covering factor
medium.

Small scale structures are also present in the atomic
ISM. Multi--epoch observations of 21\,cm absorption against
high velocity pulsars have detected opacity variations of
the ISM on a range of scales from 5\,AU to 100\,AU
(Frail et al. 1994). VLBI observations at 21\,cm in front
of 3C radio sources have shown clumps on a scale of
25\,AU (Diamond et al. 1989, Davis et al. 1996).
In the case of 21\,cm absorption against pulsars, where
small--scale opacity structures are detected towards {\it all}
line of sights, the mean opacities are between 0.1 and 2.5,
corresponding to N(HI) as low as 10$^{19}-10^{20}$\,cm$^{-2}$
(Frail et al. 1994). The opacity variations dectected are
quite high (half of them have $\delta\tau > 0.1$, Frail et al.
1994), so they cannot be accounted for by mild density
fluctuations. Clumps with density larger than
10$^5-10^6$ cm$^{-3}$ are implied (Moore \& Marscher 1995).
These clumps could represent 10--20\% of the total column
density.
Small dense clumps in the atomic gas has also been inferred
in high redshift radio galaxies (van Oijk et al. 1997), where
absorption within the Ly--$\alpha$ emission regions suggest
the presence of neutral atomic clumps of size $\sim$0.03\,pc.

\medskip

Is the molecular gas along the line--of--sight towards PKS1413+135
similar to that of our Milky Way concerning small scale structure?
Combining our molecular absorption line data with X--ray data
gives three arguments which suggest the presence of a molecular
ISM consisting of small dense clumps embedded in a more diffuse
medium:

\begin{itemize}

\item The excitation temperatures for three transitions of HCO$^+$
are consistent with a multicomponent molecular gas.
If two or more components coexist along the line--of--sight,
the assumption of a single gas component underestimates the
true column density.

\item The total column density averaged over the extent of the background
source and along the line--of--sight (atomic and molecular) is much
lower than that derived from the deficiency of low energy X--ray
photons. Since the covering factor of gas appears to be high, this
implies that we indeed underestimate the total column of gas.

\item Possible time variations in the depth of the molecular absorption
lines indirectly suggests the presence of small scale structure.

\end{itemize}

The first item has been presented in Sect.\,4.3 and 4.4. Below we
discuss the second and third item in more detail.

\subsubsection{The extinction towards PKS1413+135}

The minimum column density of molecular hydrogen towards
PKS1413+135, inferred directly from the CO(0$\rightarrow$1)
absorption, is $4 \times 10^{20}$\,cm$^{-2}$ (Sect.\,4.1).
Together with the estimated HI column density of
$1 \times 10^{21}$\,cm$^{-2}$ (where we have assumed a
covering factor of unity and a spin temperature of 100\,K),
the total column of hydrogen, $N(H) = 2N(H_2)+N(HI)$, is
$\sim 2 \times 10^{21}$\,cm$^{-2}$. This is more than
10 times lower than the column inferred from the deficiency
of low energy X--ray photons which indicates $A_v > 30$\,mag
(Stocke et al. 1992).
Is this discrepancy caused by an imprecise X--ray measurement,
by the particular distribution of the obscuring gas relative to
the radio core, or by an underestimate of the molecular and/or
atomic column density?
McHardy et al. (1994) analyzed ROSAT observations of
PKS1413+135 in the 0.4--2.4\,keV range. Although they
did not obtain enough data for a complete spectrum, the
very low flux of low--energy X--ray photons measured by
ROSAT is consistent with the deficiency observed by the
Einstein satellite (Stocke et al. 1992).
It is thus not likely that the X--ray results are erroneous.
Also, the absence of any lines from both the narrow-- and
broad emission line regions, even when viewed
in the near--IR (McHardy et al. 1994; Stocke et al. 1992;
Perlman et al. 1996), indicates that the extinction is
very high.
An obscuring dusty torus around the AGN can cause a large
extinction. In Cyg\,A, for instance, the optical extinction,
originating in a torus near the center, is estimated to be
$\sim$170\,mag, or N(H) $= 3.75 \times 10^{23}$\,cm$^{-2}$
(Ueno et al. 1994). Although 21\,cm HI absorption is detected
(Conway \& Blanco 1995), no molecular absorption is seen
despite this high column density (Drinkwater et al. 1995).
The non--detection of molecular absorption could be due to
either a lack of molecular gas as such, or that the gas is
dense enough to render the excitation temperature very high,
depopulating the lower rotational levels of existing molecules,
or that the molecular rotational levels are radiatively coupled
with the ambient radiation field (e.g. Maloney et al. 1994).
The presence of a dense torus in PKS1413+135, similar
to Cyg\,A, is, however, not likely in view of the absence of a
near--infrared excess emission, caused by dust grains heated to
high temperatures. This implies that the column density of the
molecular gas in PKS1413+135, which should be situated
at a relatively large distance from the AGN in order to
explain the lack of near--infrared emission, and 
to be compatible with the very narrow line--width,
is severly underestimated, strongly suggesting the presence
of a dense clumpy medium as emphasized in previous sections.

\subsubsection{Time variability}

Changes in the absorption lines towards background continuum
sources can be caused either by motions of small scale structure
in the absorbing gas across the angular extent of the radio
core, or by spatial changes in the radio core itself.
Such variations can give important information about both the
size of the radio core and of the scale spectrum of the molecular
interstellar medium.

The radio source PKS1413+135 consists of a compact radio core
and several emission components on a parsec--scale,
while no radio emission is seen at scales of kiloparsecs
(Perlman et al. 1994, 1996).
The parsec--scale components all have a steep--spectrum and
do not contribute to the flux at millimeter wavelengths.
The core has a spectral index $\alpha = +1.7$ between
$\lambda$18cm and $\lambda$3cm (Perlman et al. 1996).
The core remains unresolved at an angular resolution of
2.3 mas, corresponding to $\sim$7\,pc.
It is, however, likely to be significantly smaller.
Perlman et al. (1996) derive a lower limit to the size
of the core from the variability time--scale of
$\sim$10\,$\mu$arcsec, corresponding to $\sim$0.03\,pc.
The galaxy associated with PKS1413+135 is seen edge--on
(McHardy et al. 1994) and we can therefore expect that
the transverse velocity is equal to the rotational velocity.
The latter is unknown but likely to be $\sim$250\,km\,s$^{-1}$.
This corresponds to a transverse shift of 50\,AU\,yr$^{-1}$.
The time scale for a significant change of the gas 
properties due to a shift of the obscuring molecular gas
is therefore at least 100 years, possibly much longer.
The only possibility to explain much shorter variations
is the existence of high velocity ($\ga$ 25 000 km\,s$^{-1}$)
shocks propagating outwards from the radio core.

In Table\,4 we also give the continuum flux relative to that
of May 1994 (which was the highest during the extent of
these observations). The deepest absorption (largest ratio
of the first and second component) occurs when the continuum
flux is at maximum, suggesting that the changes are due to
structural changes in the background radio source
rather than transverse motion of small scale structure in
the molecular gas. Knowing the extent of the changes of the
radio core during an outburst would give us a limit to the
size of the molecular gas structures.
Future mm--VLBI observations may provide such
data. Also, the change in component 1 is likely to be
recurrent with the next outburst, if this takes place in
the same region of the radio core.

\begin{figure*}
\psfig{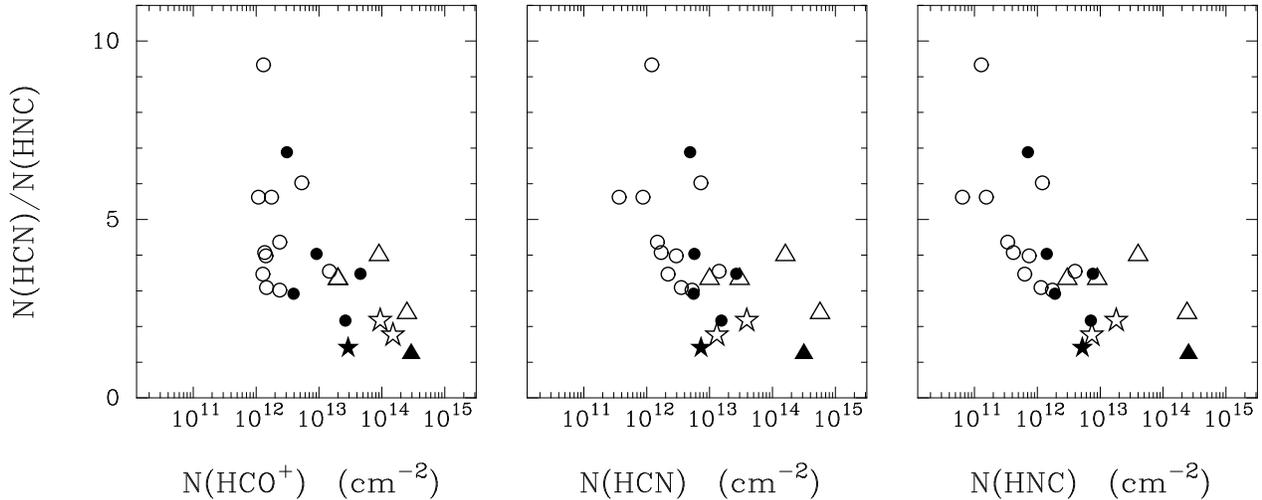}
\caption[]{The HCN/HNC ratio versus the column density of HCO$^+$,
HCN and HNC. The trend of decreasing HCN/HNC abundance ratio with
increasing column density of HCN indicated in Fig.\,11 is clearly
visible. Designations are as in Fig.\,10.}
\end{figure*}

\subsection{Molecular cloud properties}

The inferences to be made from the analysis of
multicomponent molecular gas along the line of
sight to PKS1413+135 (Sect.\,4.3 \& 4.4) is that
it is not possible to derive a correct column
density without knowledge about the structure
of the molecular ISM.
However, if we can assume homogeneity for the physical
and chemical conditions over the extent of the background
source and along the line of sigth, we can quite confidently
derive abundance ratios\footnote{Abundance is used in synonym
with column density while discussing column density
{\em ratios}.}. Furthermore, assuming similarities
between the structure of the molecular ISM in PKS1413+135
and the Milky Way, we can compare our column densities
with those derived for our Galaxy through similar observing
methods (i.e. absorption line measurements).
These assumptions are justified by the small amount of
molecular gas that actually contributes to the
observed absorption.
The angular extent of the background continuum source
is of the order 10--30\,$\mu$arcsec, as implied by time
variability (Perlman et al. 1996) and by inferences from
mm--wave VLBI of a similar radio source (Lerner et al. 1993).
The amount of molecular hydrogen probed by the line of sight
to PKS1413+135 is thus only $(0.5-2) \times 10^{-2}$\,\mo,
where we have used N(H$_2)=4 \times 10^{20}$\,cm$^{-2}$
(cf. Sect.\,4.1). The size--linewidth relation found
for molecular clouds in the Milky Way (Larson 1981;
Solomon et al. 1987) implies thay the molecular gas
is extended $\sim$1\,pc along the line of sight. Assuming
that most of the mass is contained in small clumps with
$n(H_2) = 10^4$\,cm$^{-3}$ (Sect.\,4.4), the total
H$_2$ mass is $f_V(0.4-4)$\,\mo, where $f_V$ is
the volume filling factor of the dense clumps. $f_V$ is
likely to be $<<$1.
In either case we are dealing with a very small portion
of the molecular ISM in the galaxy at z$=$0.247.
In Table\,5 we present abundance ratios for PKS1413+135.
The column densities are taken from Table\,2. The
upper and lower limits represents 1$\sigma$ limits.
Below we discuss in some more detail various results
that can be derived from these abundance ratios.

\subsubsection{Molecular oxygen O$_2$}
One of the more interesting questions in interstellar
chemistry is the abundance of O$_2$, which is supposed
to bind most of the free oxygen atoms and to be one of
the most important coolants (e.g. Goldsmith \& Langer
1978). This molecule has yet to be observed in the ISM.
The redshift of PKS1413+135 shifts the main
N(J)$=1_0\rightarrow 1_1$ line of O$_2$, which is usually
not observable from the ground, to an easily accesible
window (95.25\,GHz)
However, the rather low average opacity means that 
our limit to O$_2$/CO$<0.3 \pm 0.09$ is 20 times
higher than the upper limit derived towards B0218+357
at z$=$0.68 (Combes \& Wiklind 1995). Nevertheless,
this limit is of interest since it concerns the
ground state and thus is sensitive to excitationally
very cold gas. The transitions observed in B0218+357
were the N(J)$=1_2\rightarrow 3_2$ and the
N(J)$=1_1\rightarrow 3_2$ lines. The $1_1$ level can not
be directly populated from the ground state through
collisions, but has to be populated through
collisional excitation from N(J)$=1_0$ to $1_2$ and $3_2$
(Bergman 1995). The latter energy level corresponds
to a temperature $\sim$23\,K. Radiative transitions
from the ground state to the N(J)$=1_1$ level is allowed
and probed by the observations presented here.
This transition has now also been observed in B0218+357,
giving an improved upper limit to the O$_2$/CO ratio
for this system (Combes \& Wiklind 1997).

\begin{table}
\begin{flushleft}
\caption[]{Abundance ratios}
\small
\begin{tabular}{crclc}
\hline
 & & & & \\
${{\rm O}_2 \over {\rm CO}}$                 & $<$0.3 & $\pm$ & 0.1  &  \\
 & & & & \\
${{\rm O}_2 \over {\rm HCN}}$                & $<$0.9 & $\pm$ & 0.3  & $10^3$ \\
 & & & & \\
${^{12}{\rm CO} \over ^{13}{\rm CO}}$        & $>$74  & $\pm$ & 53   &  \\
 & & & & \\
${{\rm H}^{12}{\rm CO}^+ \over {\rm H}^{13}{\rm CO}^+}$ & $>$111  & $\pm$ & 6   &  \\
 & & & & \\
${{\rm HCN} \over {\rm CS}}$                 & $>$1.1  & $\pm$ & 0.3 &  \\
 & & & & \\
${{\rm HCN} \over {\rm H}_2{\rm CO}}$        & $>$1.5  & $\pm$ & 0.6 &  \\
 & & & & \\
${{\rm HCN} \over {\rm N}_2{\rm H}^+}$       & $>$10   & $\pm$ & 3   &  \\
 & & & & \\
\hline
 & & & & \\
${{\rm HCO}^+ \over {\rm CO}}$               & 1.3     & $\pm$ & 0.4 & $10^{-3}$ \\ 
 & & & & \\
${{\rm HCN} \over {\rm HCO}^+}$              & 0.25    & $\pm$ & 0.07 &  \\
 & & & & \\
${{\rm HCN} \over {\rm CN}}$                 & 0.4     & $\pm$ & 0.2 &  \\
 & & & & \\
${{\rm HCN} \over {\rm HNC}}$                & 1.4     & $\pm$ & 1.0 &  \\ 
 & & & & \\
\hline
\end{tabular}
\end{flushleft}
\end{table}

\subsubsection{Isotopic ratios}
The isotopic abundace ratios for CO and HCO$^+$ presented
in Table\,5 suggests relatively low abundances of the
$^{13}$C isotopic variants. The large formal error
for $^{12}$CO/$^{13}$CO comes from the assumed range
of excitation temperatures (cf. Table\,2). Since the
abundance ratio is largely independent of $T_{\rm x}$,
the true uncertainty of this ratio is much smaller.
In the ISM of the Milky Way the $^{12}$C/$^{13}$C
isotope ratio varies from $\sim$20 in the Galactic
center region to $\sim$70 in the local ISM (Wilson
\& Matteucci 1992).
The correlation between the $^{12}$C/$^{13}$C ratio and
that of the substituted isotopic molecules depends on
chemical fractionation (enhancing the isotopic variants)
and selective photoinization (decreasing the isotopic
species due to less self--shielding).
In Galactic molecular clouds as well as in nearby galaxies,
the final result is an increase in the $^{12}$CO/$^{13}$CO
ratio (and likewise for HCO$^+$) compared to the actual
$^{12}$C/$^{13}$C ratio.
The high lower limits found in PKS1413+135 therefore
suggests a low $^{13}$C abundance relative to $^{12}$C.

Whereas $^{12}$C is produced in both low and high
mass stars, as a primary product of hydrostatic
burning, $^{13}$C is only produced through incomplete
proton burning in the red giant stage of low and
intermediate mass stars (e.g. Wilson \& Matteucci
and references therein). A low $^{13}$C abundance
may therefore indicate a young chemistry, where
the low and intermediate mass stars have not yet
had time to reach the red giant stage.
$^{14}$N is produced and ejected into the ISM in much
the same way as $^{13}$C. The high lower limit
HCO$^+$/N$_2$H$^+ > 41$ found in PKS1413+135
is therefore in aggrement with the interpretation
that the low and intermediate mass stars have not
yet reached their red giant stage. In PKS1830-211
at z$=$0.89, we found a HCO$^+$/N$_2$H$^+$ ratio
of $\sim$1.4 (Wiklind \& Combes 1996a). In this
molecular absorption line system we also detect
several $^{13}$C isotopic species. Although it is
not entirely clear whether this is due to an extremely
high opacity, PKS1830--211 seems to be more chemically
evolved than PKS1413+135, despite the much higher
redshift.

\subsubsection{Dark cloud chemistry}
The equilibrium abundances in dark clouds (i.e.
molecular gas where the main ionization source is
cosmic rays) can be divided into a High and Low
Ionization Phase (HIP and LIP) (Flower et al.
1994; Le Bourlot et al. 1995; Gerin et al. 1997).
The LIP is characterized by a chemistry driven by
ion--molecule reactions where proton transfer
involving H$_{3}^{+}$ plays a major role.
The C/CO ratio is low and most of the gas phase
carbon is locked up in CO. This phase has a high
abundance of O$_2$ and molecular ions such as
HCO$^+$ and N$_2$H$^+$. Other carbon bearing
molecules, such as CN, have low abundances.
The HIP has a chemistry driven by charge transfer
reactions involving H$^+$ and is characterized by
a large C/CO ratio. This leads to low abundances
of O$_2$ and molecular ions.
Abundances of HCN and HNC are only marginally
affected by the ionization state of the gas. The
abundance ratios presented in Table\,5 mainly uses
HCN as a `reference'. 

Are our abundance ratios for PKS1413+135 consistent
with either of these two phases for dark cloud
chemistry? Comparing our abundance ratios
with the model results of Le Bourlot et al. (1995)
we find that our O$_2$/CO ratio does not constrain
the ionization state of the gas, although our limit
is of the same order as the expected O$_2$/CO ratio
for the low ionization phase. Our limit to the
O$_2$/HCN ratio is only compatible with a high
ionization phase, while our abundance ratios for
HCN/CO, HCN/HNC and the lower limit to HCN/CS only
are compatible with a low ionization phase.
Hence, the molecular absorption line data for
PKS1413+135 does not present a clear--cut case
for either of the two ionization phases.
Also, the HCO$^+$/CO ratio is too large, by more
than an order of magnitude, to be compatible
with either the high or low ionization phase.
However, as discussed by Hogerheijde et al. (1995),
turbulence might influence the chemistry,
producing enhanced amounts of HCO$^+$.
If this is the case for the high abundance of
HCO$^+$ this would indicate a diffuse gas,
since the proposed formation pathway is quenched
at $n(H_2) \ga 10^{3}$\,cm$^{-3}$.
Dark cloud chemistry shows bi--stability and can
switch from one phase to the other on a short time
scale (Le Bourlot et al. 1992, 1995), making it
possible to have a mixture of ionization phases
over small spatial scales. Improved limits on the
O$_2$ and CS abundances may enable us to put
limits to the amount of gas along the line of
sight to PKS1413+135 that can exist in HIP and
LIP states.

\subsubsection{Comparison of column densitites}
In Fig\,10 we plot the column density of HCO$^+$
versus the column densities of HCN, HNC and CS.
The plot contains data from Galactic absorption
measurements in low density gas (Lucas \& Liszt
1993, 1994, 1996) and in higher density gas seen
towards Sgr\,B2 (Greaves \& Nyman 1996), as well as
our recent results from absorption line measurements
towards Cen A (Wiklind \& Combes 1997). In addition, we
include results from two additional high--z molecular
absorption line systems 1504+377 at z$=$0.67 (Wiklind
\& Combes 1996b) and PKS1830--211 at z$=$0.89
(Wiklind \& Combes 1996a).
The dotted line is a one--to--one correspondance
between the column densities and not a fit to the
data. In Fig.\,11 we plot the column densities of HNC
and CS versus HCN in the same way as in Fig.\,10.
Despite the presence of a considerable scatter,
the most striking impression from the figures is
the remarkable correlation, over more than
3 orders of magnitude, in column density.
This suggests that the chemistry is similar for
these simple molecules in environments characterized
by very different column densities.
The high--z systems do not show any peculiarites
compared to the local values, suggesting that the
molecular ISM offers similar conditions for star
formation at earlier epochs as it does in the
present one.

The only trend in abundance ratios is found between
HCN and HNC (Fig.\,11), where the abundance of HNC
seems to decreases relative to HCN with decreasing
HCN abundance. The solid line is a least--square
fit to the HNC vs. HCN abundances:
\begin{equation}
N(HNC) = 2.3 \times 10^{-3} \left[N(HCN)\right]^{1.16}\ .
\end{equation}
In Fig.\,12 we plot the HCN/HNC abundance ratio as a function
of the HCO$^+$, HCN and HNC column densities. The HCN/HNC ratio
decreases with increasing column density in all three cases.
Isomers like HCN and HNC have similar chemistry, only differing
in a few key reactions which determine the abundance ratio.
Through chemical modelling of the HCN and HNC abundances
Schilke et al. (1992) found that the HCN/HNC ratio increases
with increasing kinetic temperature (cf. Irvine et al. 1987)
as well as with increasing molecular hydrogen density,
$n_{\rm H_2}$.
The reason for this is primarely the key reactions
\begin{eqnarray}
HNC & + & O\ \longrightarrow\ NH\  + CO \nonumber \\
\nonumber \\
HNC & + & C\ \longrightarrow\ HCN + H \nonumber
\end{eqnarray}
which are effective at relatively high temperatures
and preferentially destroy HNC relative to HCN.
A large column density does not necessarily imply a
large volume density of H$_2$ and, since the abundance
of HCN is less dependent on $T_{\rm k}$ for low H$_2$
abundances (Schilke et al. 1992), the correlations
between the HCN/HNC and column densitites seen in Fig.\,12
is likely to be a temperature effect, affecting primarily
the HNC abundance. The lowest column densities being
associated with the kinetically warmest gas.
However, the dependence of the HCN/HNC abundance ratio
on {\em both} temperature and density makes it less valuable
as a temperature probe then previously thought.

The high--z systems all tend to have HCN/HNC ratios $\sim$2,
among the lowest of the line--of--sights presented in
Figs.\,10--12. It is unclear whether this is a selection
bias towards both kinetically and excitationally cold gas,
or an effect of low metallicity, making the HNC depletion
reactions less effective.

\begin{table}
\begin{flushleft}
\caption[]{Redshifts and variations of molecular mass}
\scriptsize
\begin{tabular}{lcc}
\hline
 & & \\
\multicolumn{1}{c}{Transition}                   &
\multicolumn{1}{c}{z$_a$}                        &
\multicolumn{1}{c}{$\left[{\Delta{\rm z} \over {\rm (1+z)}}\right]^{a)}$} \\
 & & \\
\hline
 & & \\
{\bf PKS1413+135} \\
 & & \\
HI                       & $0.24671 \pm 1 \times 10^{-5}$   &
---                    \\
 & & \\
CO(0$\rightarrow$1)      & $0.2467091 \pm 3 \times 10^{-7}$ &
$(+0.7 \pm 8.0) \times 10^{-6}$  \\
 & & \\
HCN(1$\rightarrow$2)     & $0.2467112 \pm 1 \times 10^{-6}$ &
$(-0.9 \pm 8.1) \times 10^{-6}$ \\
HCN(2$\rightarrow$3)     & $0.2467105 \pm 3 \times 10^{-7}$ &
$(-0.4 \pm 8.0) \times 10^{-6}$ \\
 & & \\
HCO$^+$(1$\rightarrow$2) & $0.2467102 \pm 3 \times 10^{-7}$ &
$(-0.2 \pm 8.0) \times 10^{-6}$ \\
HCO$^+$(2$\rightarrow$3) & $0.2467097 \pm 3 \times 10^{-7}$ &
$(-0.2 \pm 8.0) \times 10^{-6}$  \\
 & & \\
HNC(1$\rightarrow$2)     & $0.2467114 \pm 3 \times 10^{-7}$ &
$(-1.1 \pm 8.0) \times 10^{-6}$ \\
 & & \\
\hline
 & & \\
{\bf B0218+357} \\
 & & \\
HI                       & $0.68466 \pm 4 \times 10^{-5}$   &
---                              \\
 & & \\
$^{13}$CO(1$\rightarrow$2) & $0.684693 \pm 1 \times 10^{-6}$ &
$(+2.0 \pm 2.4) \times 10^{-5}$ \\
 & & \\
\hline
 & & \\
{\bf B3\,1504+377} \\
 & & \\
HI                       & $0.67324 \pm 1 \times 10^{-5}$   &
--- \\
                         & $0.67343 \pm 1 \times 10^{-5}$   &
--- \\
 & & \\
HCO$^+$(1$\rightarrow$2) & $0.673184 \pm 1 \times 10^{-6}$   &
$(-3.4 \pm 0.6) \times 10^{-5}$ \\
                         & $0.673327 \pm 1 \times 10^{-6}$   &
$(-6.2 \pm 0.6) \times 10^{-5}$ \\
 & & \\
\hline
 & & \\
{\bf PKS1830} \\
 & & \\
HI                       & ---                              &
--- \\
 & & \\
HCO$^+$(1$\rightarrow$2)        & $0.8858261 \pm 3 \times 10^{-7}$  &
 \\
H$^{13}$CO$^+$(1$\rightarrow$2) & $0.8858647 \pm 6 \times 10^{-7}$ &
 \\
HCO$^+$(2$\rightarrow$3)        & $0.8858298 \pm 3 \times 10^{-7}$ &
 \\
N$_2$H$^+$(1$\rightarrow$2)     & $0.8858262 \pm 4 \times 10^{-7}$ &
 \\
H$_2$CO(1$_{10}\rightarrow 2_{11}$) & $0.8858263 \pm 4 \times 10^{-7}$ &
 \\
CS(2$\rightarrow$3)             & $0.8858295 \pm 3 \times 10^{-7}$ &
 \\
 & & \\
\hline
\end{tabular}
\ \\
a)\ $\Delta z/(1+z) = (z_{HI} - z_{mol})/(1+z_{HI})$ \\
HI references: PKS1413+135 (Carilli et al. 1992), B0218+357 (Carilli
et al. 1993), B3\,1504+377 (Carilli et al. 1996).  \\
Molecular line references: PKS1413+135 (this paper), B0218+357 (Combes
\& Wiklind 1995), B3\,1507+377 (Wiklind \& Combes 1996b), PKS1830--211
(Wiklind \& Combes 1996a,1997).
\end{flushleft}
\end{table}

\begin{figure}
\psfig{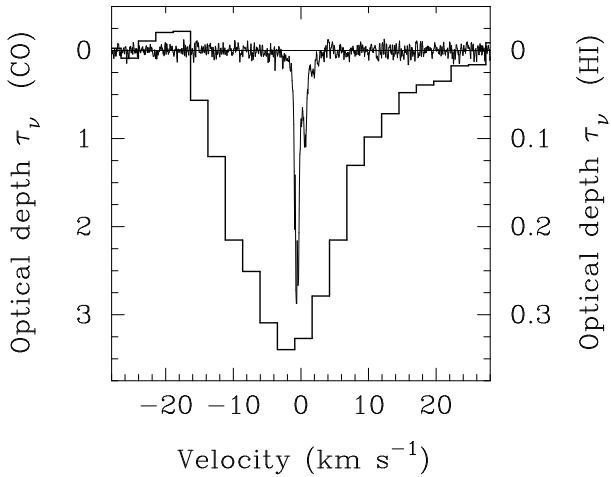}
\caption[]{A comparison of the CO(0$\rightarrow$1) and 21\,cm HI absorption
towards PKS1413+135. In both cases the optical depth is plotted. Notice that
Fig.\,3 in Wiklind \& Combes 1994 suffers from an erroneous velocity
scale, causing the CO and HI lines to be separated by $\sim$11\,km\,s$^{-1}$.}
\end{figure}

\subsection{Invariance of physical constants}

A potential diagnostic application of molecular
absorption lines at high redshift is to check
the invariance of fundamental physical parameters.
(cf. Thompson 1975).
The energy difference between two adjacent
rotational levels in a molecule consisting of
two atoms is proportional to $(\mu r^2)^{-1}$,
where $r$ is the bond length and $\mu$ is the
reduced mass of the atoms. A vibrational transition
of the same molecule has, to a first order approximation,
a $\mu^{1/2}$ dependence on the reduced mass.
Comparison of a vibrational--rotational spectrum
from an earlier epoch with a present one can
therefore be used to set limits on the changes
of the proton and neutron masses. This has been
done for molecular hydrogen seen in absorption in
a damped Lyman--$\alpha$ system at z$=$2.81
(Foltz et al. 1988, Varshalovich \& Potekhin 1996),
giving a limit of $2 \times 10^{-4}$ to $\Delta m_p/m_p$.
For pure rotational transitions one can compare
the line frequencies (or similarly, the measured
redshifts) with the frequencies of electronic
transitions. This gives a measure of the invariance
of the electron to proton mass. It is also possible
to compare the 21\,cm HI line with a molecular
rotational line. The formers energy difference
to first order depending on $m_{e}^2$, again
giving a measure of the invariance of the
$m_e/m_p$ ratio (see also Tubbs \& Wolfe 1980 for
a comparison of the 21\,cm line with optical resonance
transitions).

In an Appendix we express the energy levels of molecular
rotational lines in fundamental physical parameters. It is
shown that for a differential measurement, i.e. measuring the
redshifts of two different molecular species, $\Delta z/(1+z)$
is directly proportional to $\Delta \mu/\mu$. Strictly speaking,
this is only valid when the molecule is approximated as a rigid
rotator, but this is a good approximation for low lying $J$ states.
In Table\,6 we list the observed redshifts of the molecular
absorption lines in PKS1413+135, together with some lines from
B\,0218+357, B3\,1504+377 and PKS1830--211. We also list the
redshifts derived from 21\,cm HI absorption when available.
The redshifts have been obtained using
Gaussian fits. We can compare both the redshifts derived for different
molecules and the redshifts of the HI and
molecular lines. If changes in the electron mass are neglible, the
$\Delta z/(1+z)$ measures $\Delta \mu/\mu$, to first order,
in both cases. The values given in Table\,6 are the difference
between the 21\,cm HI and molecular redshifts.
In Figure\,13 we show the 21\,cm HI absorption (Carilli et al.
1992) together with our CO(0$\rightarrow$1) absorption towards
PKS1413+135. It is quite clear that the HI spectra
limit the accuracy with which $\Delta z$ can be measured.
Except for B3\,1504+377, all the values in Table\,6 are compatible
with $\Delta z = 0$. For the 21\,cm HI absorption obtained towards
B3\,1504+377 (Carilli et al. 1997), the discrepancy 
between HI and molecular velocities is of the order of 30\,km\,s$^{-1}$, 
i.e. $\Delta z \approx 10^{-4}$, an order of magnitude larger
than the HI accuracy. The HI result has been confirmed to an accuracy
better than 1\,km\,s$^{-1}$ with the Westerbork Synthesis Radio Telescope
(Carilli et al 1997, preprint). Hence we reach here the basic
limitation of the method, consisting of observing different
atoms or molecules. It cannot be expected that all lines
peak at the same velocity, since their local association is not
perfect, and their excitation might be different. Supporting this, note
that the molecular component in absorption in B3\,1504+377
at $z = 0.67150$ has no HI counterpart.

To obtain more statistics, atomic and molecular absorptions in the
Milky Way must be considered. 
Liszt \& Lucas (1997) compared the absorption profiles of
several different molecular lines with that of 21\,cm HI and
found that the velocity of the peak opacity often differs
between the molecular and atomic lines. This shows that
the absorptions do not arise in the same gas and any comparison
is limited by the possible velocity difference of the two
ISM components where the absorptions originates.

It is also possible to use different transitions from the same
molecule to estimate $\Delta \mu/\mu$; While the rotational
constant is proportional to $\mu^{-1}$, the centrifugal
distortion coefficient is proportional to $\mu^{-2}$
(Townes \& Schawlow 1975). In the Appendix we show that we
actually have a dependence on the invariance of the electron
mass and the fine structure constant in this case.
Comparing the J$=$1$\rightarrow$2 and J$=$2$\rightarrow$3
lines of HCO$^+$ in PKS1413+135 we get at the
1$\sigma$ level:
$\Delta z/(1+z) = (4.0 \pm 3.4) \times 10^{-7}$.
Similarly for HCN:
$\Delta z/(1+z) = (5.6 \pm 8.4) \times 10^{-7}$.
Comparing CO(0$\rightarrow$1) and HCO$^+$(1$\rightarrow$2)
we get: $(-1.1 \pm 0.3) \times 10^{-6}$. This last result
is marginally different from zero at a 3$\sigma$ level.
However, even when comparing molecular transitions one
has to face the possibility that the lines do not come
from the same location in phase space.
HCO$^+$, for instance, has been shown to exist in regions
where the CO molecular is no longer self--shielded (Lucas
\& Liszt 1996), and its absorption profile can therefore be
weighted with velocity components not seen in the CO line.
A further complication is that HCO$^+$ is an ion and susceptible
to magnetic fields present in the ISM. This can give it
a drift velocity different from neutral molecular species.
It is possible to have absorption lines dominated by different
velocity components even when considering transitions from
the same molecular species. This is due to the nonlinear
dependence of the opacity to the excitation temperature
(Sect.\,4.2). Different transitions can therefore be weigthed
differently by gas along the line of sight.

To summarize, measurements of redshift differences between
molecular absorption lines and 21\,cm HI are likely to
be dominated by velocity differences between the two gas
components along the line of sight. This is a severe concern
when comparing different molecular species and even different
transitions of the same molecule.
Taken at face value, the HI accuracy limits the nucleon
mass variation to typically $\Delta \mu/\mu \la 10^{-5}$
at a 3$\sigma$ level. In one source (B3\,1504+377) a discrepancy
as large as $10^{-4}$ has been observed.

\section{Summary}

We present a multi--line study of an absorbing
molecular cloud in the edge--on galaxy hosting the
radio--source PKS1413+135, at z$=$0.24671. We have
shown  that the total column densities obtained are
lower limits, since the absorption essentially
traces the diffuse extended molecular gas, and is
only weakly weighted towards the dense phase which
provides most of the mass. The size of the absorbing
structure along the line of sight must be of parsec
scale, given its narrow total width.
This situation is favorable to detect time
variations in the absorption profile, if there exists knots
in the radio source moving close to the velocity of light.
Variations on time--scales of a month then correspond
to structures of $> 10^{3}$\,AU. We report possible
time variation, affecting the relative ratio
of two absorbing components, and therefore the
shape of the profile. This is to a large extent
independent of the  continuum level and calibration
uncertainties. Such time--variations are not surprising,
in view of the small scale structure that has been
detected in the ISM of the Milky Way.

Because of the high spectral precision of the
millimeter techniques, it has been proposed that observation
of molecules at high redshift could constrain variations
of fundamental physical constants, such as the mass of
the molecules. However, different molecules or different
kind of lines (rotational, vibrational, etc..) should then
be compared. Concerning the comparison between HI and molecular lines,
it was shown that the HI accuracy limits the nucleon mass variation
to $\Delta \mu/\mu \la 10^{-5}$ at 3$\sigma$ in many cases, but
a discrepancy as large as $10^{-4}$ has been observed in one of
the sources. Excitation problems in the not
well--known multiphase structure of the medium prevents
reaching conclusive results.

\acknowledgements

We are very grateful to the IRAM engineers and operators for the
impressive work in tuning dizains of new frequencies during several
observing runs and the generous allocation of observing time by IRAM.
TW acknowledges support from NFR (the Swedish Natural Science Research
Council) for this research. 


\section{Appendix}

Here we present in more detail the use of molecular rotational
lines as a probe of the invariance of physical constants over
large spatial and temporal scales. In particular, we show that
there is a limit to the accuracy of molecular rotational
lines, coming from our incomplete solution of the Schr\"{o}dinger
wavefunction for a complex system such as a molecule.

\subsection{A diatomic rigid rotator}

As a first approximation we will look at a diatomic molecule without
considering the centrifugal distortion and the zero--point vibrational
energy.

The energy of a rotational level $J$ is
\begin{equation}
E_{\rm rot} = {\hbar \over 4\pi I} J(J+1)\ ,
\end{equation}
where $J$ is the angular momentum quantum number and $I$ is the molecular
moment of inertia about axes perpendicular to the internuclear axis
(cf. Townes \& Schawlow 1975). For a diatomic molecule the moment of
inertia can be expressed as $\mu r_{\rm e}^{2}$, where $\mu$ is the
reduced mass of the two atoms and $r_{\rm e}$ is the equilibrium
intranuclear distance of the two atoms.
To within a factor of 2--3, we can set $r_{\rm e} \approx a_0$, the
Bohr radius. Expressed in fundamental parameters\footnote{These are
usually called `constants', but since we are testing their invariance,
we will refer to them as `parameters'.}, the energy of a rotational
level $J$ is
\begin{equation}
E_{\rm rot} \approx \left({c^2 \over 2 \hbar}\right)
{m_{\rm e}^{2} \alpha^{2} \over \mu} J(J+1)\ ,
\end{equation}
where $m_{\rm e}$ is the electron mass and $\alpha$ the fine structure
constant. The frequency of a rotational transition following the
selection rule $\Delta J = \pm 1$ is
\begin{equation}
\nu = {2 \hbar \over 4\pi I}(J+1) \approx
\left({c^2 \over h}\right) {m_{\rm e}^{2} \alpha^2 \over \mu} (J+1)\ .
\end{equation}
This gives the correct frequency of a transition to within a factor
of a few. The discrepancy arises from the approximation of
the effective internuclear separation as the Bohr radius, but does
not affect the analysis of the fundamental physical parameters.
A shift in the line frequency due to changes in the reduced mass $\mu$,
the electron mass $m_{\rm e}$ and the fine structure constant $\alpha$
is therefore
\begin{equation}
{\Delta \nu \over \nu} = \sqrt{\left({\Delta \mu \over \mu}\right)^2 +
4\left({\Delta m_{\rm e} \over m_{\rm e}}\right)^2 +
4\left({\Delta \alpha \over \alpha}\right)^2}\ .
\end{equation}
If the redshift of an object causing the observed transition
is known apriori, we can directly relate the shift in frequency
or, similarly, in velocity $\Delta v = c \Delta \nu/\nu$ to
combined changes in the three parameters $\mu, m_{\rm e}$ and
$\alpha$. In order to measure a shift of 1\,km\,s$^{-1}$, we
must have $\Delta \mu/\mu \sim 3 \times 10^{-6}$
(if $\Delta m_{\rm e} = \Delta \alpha = 0$). 
However, any determination of a redshift involves a measurement
of line radiation, with the possibility of being affected by
changing physical parameters (not necessarily the same as for
molecular rotational transitions). Hence, we must make a
`differential' measurement in order to test their invariance.
In the case of a rigid diatomic molecule, this is done
by comparing two different species. What we measure is two ratios
of frequencies; the `laboratory' (called $lab$) and the observed
frequency (called $obs$):
\begin{equation}
{\nu_{\rm lab} \over \nu_{\rm obs}} = 1 + z\ .
\end{equation}
For two different molecules $a$ and $b$, we can now write:
\begin{equation}
1 - {\left[{\nu_{\rm lab} \over \nu_{\rm obs}}\right]_{\rm b} \over
\left[{\nu_{\rm lab} \over \nu_{\rm obs}}\right]_{\rm a}} =
{\Delta z \over 1 + z_{\rm a}}\ .
\end{equation}
Since we are comparing two molecular species at two different epochs,
the electron mass and the fine structure constant will drop out and
we have
\begin{eqnarray}
{\Delta z \over 1 + z_{\rm a}} & = & 
{\Delta R_{\mu} \over R_{\mu_0}} \\
\nonumber \\
\Delta R_{\mu} & = & \left({\mu_{\rm b} \over \mu_{\rm a}}\right)_{0}
- \left({\mu_{\rm b} \over \mu_{\rm a}}\right)_{z} \\
\nonumber \\
R_{\mu_0} & = & \left({\mu_{\rm b} \over \mu_{\rm a}}\right)_{0} \ .
\end{eqnarray}
In Eqs.\,20--22 we use $0$ as a subscript for mass ratios obtained in
the laboratory (z$=$0) and $z$ for mass ratios at a distant epoch.
Comparison of molecular rotational transitions are thus sensitive to
the strong force only, affecting the proton and neutron masses. For
practical purposes, the largest effect is obtained when $R_{\mu_0}$ 
is small, i.e. $\mu_{\rm a} >> \mu_{\rm b}$.

\subsection{Non--rigid diatomic molecules}

The next step in the complexity of a diatomic molecule is the
centifugal distortion and the zero--point vibrational energy.
As the rotational quantum number $J$ increases the molecule
rotates faster and the internuclear separation increases.
This leads to a larger moment of inertia and a decrease in
the energy compared with the rigid rotator. The zero--point
vibrational energy introduces a periodic oscillation in the
internuclear separation, affecting the moment of inertia and
the energy levels.

A separate treatment of the atoms and the electrons (the
Born--Oppenheimer approximation) is necessary in order to
describe the total potential in such a way that the Schr\"{o}dinger
wave function can be solved. Ultimately, as we will discuss
in the next subsection, the Born--Oppenheimer approximation
puts a limit to the accuracy with which molecular rotational
transitions can be used to test the invariance of physical
parameters.

Dunham (1932, Phys Rev 41, 721) gave the vibrational and rotational
energy levels for any potential which can be expanded in powers of 
$(r-r_{\rm e})$ (see also Townes \& Schawlow 1975).
However, we will use the simpler Morse potential in order to emphasize
the dependence on fundamental physical constants. The Morse potential
is written
\begin{equation}
U(r) = D(1-\exp(-a(r-r_{\rm e}))^2\ ,
\end{equation}
where $D$ is the dissociation energy of the molecule, $r_{\rm e}$
is the equilibrium intranuclear separation and $a$ is a constant.
The frequency of a rotational transition ($v=0$) becomes
\begin{eqnarray}
\nu & = & 2B_e(J+1) - 4D_e(J+1) - \alpha_e(J+1) \\
\nonumber \\
B_e & = & {\hbar \over 4\pi I_e} \approx
\left({c^2 \over 2h}\right) {m_{\rm e}^{2} \alpha^2 \over \mu} \\
\nonumber \\
D_e & = & {8\pi^2 \mu B_e^3 \over a^2 D} \approx
\left({\pi^2 c^6 \over a^2 D h^3}\right)
\left({m_{\rm e}^{3}\alpha^3 \over \mu}\right)^2 \\
\nonumber \\
\alpha_e & = & \left({3 c^3 \over \sqrt{32D}\pi \hbar}\right)
\left({m_{\rm e} \alpha \over \mu^{1/2}}\right)^3
\left[1 - {\hbar \over a c m_{\rm e} \alpha}\right] \\ .\
\end{eqnarray}
If we for simplicity disregard the zero--point vibrational energy, we
can write the line frequency as
\begin{eqnarray}
\nu & = & \left({c^2 \over h}\right) {m_{\rm e}^{2} \alpha^2 \over \mu}
(J+1) \left[1 - \left({c^3 \over a^2 D \hbar^2}\right)
{m_{\rm e}^{3} \alpha^3 \over \mu} (J+1)^2\right]
\end{eqnarray}
The first term is identical to that obtained for the rigid rotator and
the second term is the effect of centrifugal stretching (lowering the
frequencies). Due to this term, the dependence on $m_{\rm e}$ and
$\alpha$ at $z=0$ and at $z$ does not cancel when comparing the
redshifts for two molecules $a$ and $b$, as was the case for the
rigid rotator. Hence, $\Delta z/(1+z)$ depends on all three parameters;
$\mu$, $m_{\rm e}$ and $\alpha$. For low $J$--values, however, the
dependence on the electron mass and the fine structure constant is
small, and the rigid rotator represents a good first order approximation.
The nonlinear dependence on $J$ in the centrifugal distortion means that
it is possible to compare the redshifts derived from different rotational
transitions of the same molecule.

\subsection{Limitations to the accuracy}

As discussed in Sect.\,5.4, the accuracy of using comparative
measures of redshifts derived from either different emission
mechanisms (optical, 21\,cm hyperfine line, molecular rotational
lines, etc) is limited by velocity gradients in the emitting
medium and the fact that different gas masses contribute
differently to the various emission/absorption processes.
This is a grave concern also when using molecular rotational
transitions, even when the same molecular species is used,
but with different $J$ levels. However, even if this was not
the case, the accuracy is limited by our inexact quantum
mechanical description of a molecule.

There are two main uncertainties which affect a test of the
invariance of fundamental physical parameters. The first is
anharmonicity of the potential. This is particularly true
for the Morse potential used here. The situation is improved
somewhat by using the method of Dunham (cf. Townes \& Schawlow
1975), but the method involves parameters which can be calculated
from spectroscopically observable quantities and they ultimately
put a limit to how accurate the ratio of two reduced masses
can be calculated to $\sim 10^{-7}$ (Townes \& Schawlow 1975).
The second uncertainty involves the Born--Oppenheimer approximation,
where the electronic and atomic contributions to the total
potential are treated separately. Although electrons can be
considered to be more or less spherically distributed around
their respective nuclei, and thus not contributing to the moment
of inertia (the so called `slip effect'), electrons in the
valence shell will give a contribution to the moment of
inertia of $\sim n_{\rm e}m_{\rm e}r^2$, where $n_{\rm e}$
is the number of electrons in the valence shell, $m_{\rm e}$
is the electron mass and $r$ is an average radius from the
nucleus with which the valence shell is associated. This
contribution to the moment of inertia is very small indeed,
but when deriving mass ratios it limits the accuracy to
$\sim 10^{-6}$ (smaller for heavy molecules and larger for
lighter ones).
An effect which is closely related to the valence electrons
is so called L--uncoupling. Rotation tends to excite the
valence electrons from their ground $^{1}\Sigma$ state with
zero angular momentum, to excited $^{1}\Pi$ states, with unit
angular momentum. This in itself affects the rotational energy,
but a larger effect comes from the fact that the $\Pi$ state
also produces a large magnetic field at the positions of the
nuclei. This induces magnetic hyperfine interaction, again
changing the rotational energy levels.
This L--uncoupling can be measured in the laboratory and is
found to introduce uncertainties of $\sim 10^{-6}$ in estimated
mass ratios (Townes \& Schawlow 1975).

Hence, the uncertainites associated with estimates of mass
ratios of molecules is of the order $10^{-6}$. This value
is of the same order as the $\Delta \mu/\mu$ necessary to
produce a shift of $\sim$1\,km\,s$^{-1}$ relative to the
`true' redshift of a molecular rotational line. Smaller
shifts than this will therefore not be useful in testing
the invariance of physical parameters.

\end{document}

%% file: alias.tex
\newcommand{\mo}    {\hbox{${\rm M}_{\odot}$}}

\newcommand{\lo}    {\hbox{${\rm L}_{\odot}$}}

\newcommand{\kms}   {\hbox{${\rm km\,s}^{-1}$}}
\newcommand{\ms}    {\hbox{${\rm m\,s}^{-1}$}}
\newcommand{\kmsmpc}{\hbox{${\rm km\,s}^{-1}{\rm Mpc}^{-1}$}}

\newcommand{\asec}  {\hbox{$^{\prime\prime}$}}

\newcommand{\tastar}{\hbox{$T_{\rm A}^{*}$}}
\newcommand{\cmsq}  {\hbox{${\rm cm}^{-2}$}}

\newcommand{\hI}    {\hbox{${\rm HI}$}}


%% file: 6227.bbl
\begin{thebibliography}{}

\bibitem{} Bergman P., 1995, ApJ 445, L167

\bibitem{} Bregman J.N., Lebofsky M.J., Aller M.F., et al., 1981,
Nature 293, 714

\bibitem{} Briggs F., 1988, in QSO Absorption Lines, eds. J.C.
Blades, D. Turnshek, C. Norman, Cambridge University Press, p.\,275

\bibitem{} Carilli C.L., 1995, Journal of Astrophysics and Astronomy, Vol. 
16 (Supplement), Proceedings of the Sixth IAU Asian-Pacific Regional Meeting,
eds. V. Kapahi N. Dadhich, G. Swarup, and J. Narlikar, p. 163

\bibitem{} Carilli C.L., Menten K.M., Reid M.J., Rupen M.P., 1997,
ApJ L89

\bibitem{} Carilli C.L., Rupen M.P., Yanny B., 1993, ApJ 412, L59

\bibitem{} Carilli C.L., Perlman E.S., Stocke J.T., 1992, ApJ 400, L13

\bibitem{} Combes F., Wiklind T., 1996, in Cold Gas at High Redshift,
eds. M.N. Bremer, P. van der Werf, H.J.A. R\"{o}ttgering, C.L. Carilli,
Kluwer Academic Pub., p.\,215

\bibitem{} Combes F., Wiklind T., 1995, A\&A 303, L61    

\bibitem{} Conway J.E., Blanco P.R., 1995, ApJ 449, L131

\bibitem{} Davis, R.J., Diamond P.J., Goss W.M., 1996, MNRAS 283, 1105

\bibitem{} Diamond P.J., Goss W.M., Romney J.D. et al., 1989, ApJ 347, 302

\bibitem{} Drinkwater M., Combes F., Wiklind T., 1996, A\&A 312, 771

\bibitem{} Falgarone E., Phillips T.G., 1996, ApJ 472, 191

\bibitem{} Falgarone E., Puget J-L., P\'erault M., 1992, A\&A 257, 715

\bibitem{} Flower D.R., Le Bourlot J., Pineau des For\^{e}ts G., Roueff E.,
1994, A\&A 282, 225

\bibitem{} Foltz C.B., Chaffee F.H., Black J.H., 1988, ApJ 324, 267

\bibitem{} Frail D.A., Weisberg J.M., Cordes J.M., Mathers C., 1994, ApJ 436, 144

\bibitem{} Gerin M., Falgarone E., Joulain K., et al., 1997, A\&A 318, 579

\bibitem{} Goldsmith P.F., Langer W.D., 1978, ApJ 222, 881

\bibitem{} Greaves J.S., Nyman L.--\AA., 1996, A\&A 305, 950

\bibitem{} Hogerheijde M.R., de Geus E.J., Spaans M., van Langevelde H.,
1995, ApJ 441, L93

\bibitem{} Irvine W.M., Goldsmith P.F., Hjalmarson \AA., 1987, in
Interstellar Processes, eds. D.J. Hollenbach, H.A. Thronson, Reidel
Publ. Co., p.\,561

\bibitem{} Larson R.B., 1981, MNRAS 194, 809

\bibitem{} Le Bourlot J., Pineau des For\^{e}ts G., Roueff E., 1995,
A\&A 297, 251

\bibitem{} Le Bourlot J., Pineau des For\^{e}ts G., Roueff E., 1993,
ApJ 416, L87

\bibitem{} Lerner M., B{\aa}{\aa}th L., Inoue M., et al.,
1993, A\&A 280, 117

\bibitem{} Liszt H., Lucas R., 1997 A\&A, in press

\bibitem{} Liszt H., Lucas R., 1996 A\&A, 314, 917

\bibitem{} Lovas F.J., 1992, J. Phys. Chem. Ref. Data 21, 181

\bibitem{} Lucas R., Liszt H., 1996, A\&A 307, 237

\bibitem{} Lucas R., Liszt H., 1994, A\&A 282, L5

\bibitem{} Lucas R., Liszt H., 1993, A\&A 276, L33

\bibitem{} Maloney P.R., Begelman M.C., Rees M.J., 1994, ApJ 432, 606

\bibitem{} Marscher A.P., Moore E.M., Bania T.M., 1993, ApJ 419, L101

\bibitem{} Marscher A.P., Stone A.L., 1994, ApJ 433, 705

\bibitem{} McHardy I.M., Merrifield M.R., Abraham R.G., Crawford C.S.,
1994, MNRAS 268, 681

\bibitem{} McHardy I.M., Abraham R.G., Crawford C.S., et al.,
1991, MNRAS 249, 742

\bibitem{} Moore E.M., Marscher A.P., 1995, ApJ 452, 671

\bibitem{} Perlman E.S., Carilli C.L., Stocke J.T., Conway J., 1996,
AJ 111, 1839

\bibitem{} Perlman E.S., Stocke J.T., Shaffer D.E., Carilli C.L.,
Ma C., 1994, ApJ 424, L69

\bibitem{} Schilke P., Walmsley C.M., Pineau des Forets, et al.,
1992, A\&A 256, 595

\bibitem{} Solomon P.M., Rivolo A.R., Barrett J., Yahil A., 1987, ApJ
319, 730

\bibitem{} Stocke J.T., Wurtz R., Wang Q., Elston R., Jannuzi B.T.,
1992, ApJ 400, L17

\bibitem{} Thompson R.I., 1975, Ap Letters 16, 3

\bibitem{} Townes C.H., Schawlow A.L., 1975, Microwave Spectroscopy,
Dover Pub. New York

\bibitem{} Tubbs A.D., Wolfe A.M., 1980, ApJ 236, L105

\bibitem{} Ueno S., Koyama K., Nishida H., Yamauchi S., Ward M.J., 1994,
ApJ 431, L1

\bibitem{} van Dishoeck E.F., Black J.H., 1987, Physical Processes in
Interstellar Clouds, eds. G.E. Morfill, M. Scholer, Reidel Publishing
Company, p.\,241

\bibitem{} van Oijk R., R\"{o}ttgering H.J.A., Miley G.K., Hunstead R.W.,
1997, A\&A 317, 358

\bibitem{} Varshalovich D.A., Potekhin A.Y., 1996, Astron Letters 22, 1

\bibitem{} Wiklind T., Combes F., 1997, A\&A in press       

\bibitem{} Wiklind T., Combes F., 1996b, A\&A 315, 86       

\bibitem{} Wiklind T., Combes F., 1996a, Nature 379, 139    

\bibitem{} Wiklind T., Combes F., 1995, A\&A 299, 382       

\bibitem{} Wiklind T., Combes F., 1994b, A\&A 288, L41      

\bibitem{} Wiklind T., Combes F., 1994a, A\&A 286, L9       

\bibitem{} Wilson T.M., Matteucci F., 1992, A\&AR 4, 1

\end{thebibliography}
